\documentclass[aoas]{imsart}

\RequirePackage{amsthm,amsmath,amsfonts,amssymb}
\RequirePackage[authoryear]{natbib}
\RequirePackage[colorlinks,citecolor=blue,urlcolor=blue]{hyperref}
\RequirePackage{graphicx}
\RequirePackage[caption=false]{subfig}
\RequirePackage{bbm}
\usepackage{chngcntr}
\usepackage{floatrow}
\startlocaldefs
\numberwithin{equation}{section}

\DeclareMathOperator*{\argmax}{arg\,max}
\DeclareMathOperator*{\argmin}{arg\,min}

\endlocaldefs

\begin{document}

\begin{frontmatter}
\title{Learning Risk Preferences in Markov Decision Processes: an Application to the Fourth Down Decision in the National Football League}
\runtitle{Learning Risk Preferences in Markov Decision Processes}

\begin{aug}
\author[A]{\fnms{Nathan}~\snm{Sandholtz}\ead[label=e1]{nsandholtz@stat.byu.edu}},
\author[B]{\fnms{Lucas}~\snm{Wu}\ead[label=e2]{lwu@zelusanalytics.com}},
\author[C]{\fnms{Martin}~\snm{Puterman}\ead[label=e3]{martin.puterman@sauder.ubc.ca}},
\and
\author[D]{\fnms{Timothy C.Y.}~\snm{Chan}\ead[label=e4]{tcy.chan@utoronto.ca}}
\address[A]{Department of Statistics,
Brigham Young University\printead[presep={,\ }]{e1}}

\address[B]{Zelus Analytics\printead[presep={,\ }]{e2}}

\address[C]{Sauder School of Business,
University of British Columbia\printead[presep={,\ }]{e3}}

\address[D]{Department of Mechanical \& Industrial Engineering,
University of Toronto\printead[presep={,\ }]{e4}}
\end{aug}

\begin{abstract}
For decades, National Football League (NFL) coaches' observed fourth down decisions have been largely inconsistent with prescriptions based on statistical models.  In this paper, we develop a framework to explain this discrepancy using an inverse optimization approach.  We model the fourth down decision and the subsequent sequence of plays in a game as a Markov decision process (MDP), the dynamics of which we estimate from NFL play-by-play data from the 2014 through 2022 seasons.  We assume that coaches' observed decisions are optimal but that the risk preferences governing their decisions are unknown.  This yields an inverse decision problem for which the optimality criterion, or risk measure, of the MDP is the estimand.  Using the quantile function to parameterize risk, we estimate which quantile-optimal policy yields the coaches' observed decisions as minimally suboptimal.  In general, we find that coaches' fourth-down behavior is consistent with optimizing low quantiles of the next-state value distribution, which corresponds to conservative risk preferences.  We also find that coaches exhibit higher risk tolerances when making decisions in the opponent's half of the field as opposed to their own half, and that league average fourth down risk tolerances have increased over time.
\end{abstract}

\begin{keyword}
\kwd{Inverse optimization}
\kwd{MDP}
\kwd{quantile function}
\kwd{decision analysis}
\kwd{sport}
\end{keyword}

\end{frontmatter}



\section{Introduction}

The fourth down decision in American football is one of the most well-studied and popularly discussed decision problems in sports analytics.  In the National Football League (NFL), each team has four ``downs" (i.e., chances) to advance the football 10 yards.  On fourth down, coaches face a decision involving a high-risk, high-reward trade-off: they can either ``go for it", attempting to gain the remainder of the 10 yards and thus a first down, or they can kick the ball either by attempting a field goal (when the team is close enough to the opponent's end zone) or by punting (when the team is farther away). Punting concedes possession of the ball, but it does so by putting the other team in a worse field position, typically deeper in their own half of the field.  Attempting a field goal can yield an immediate three points if successful, but a miss gives the opposing team possession of the ball from the location of the kick.\footnote{If the field goal is taken on or inside the 20-yard line, the ball is placed at the 20-yard line after a missed field goal attempt.} Going for it is generally considered the more risky option since a failed attempt means losing possession in a more favorable position for the opponent (if punting is the alternative decision) or losing out on three points (if attempting a field goal is the alternative decision).

\begin{figure}
\centering
    \subfloat[]{
        \includegraphics[width=.357\linewidth]{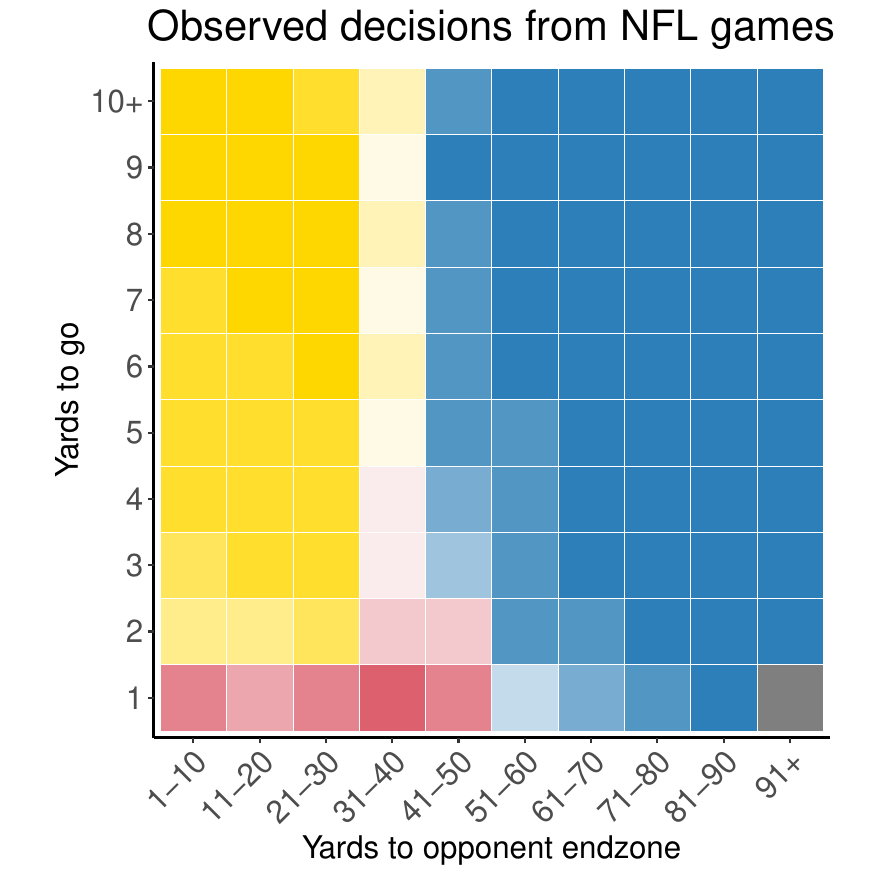}
        \label{fig:observed_dec}
    }
    \hspace{2em}%
    \subfloat[\quad\quad\quad\quad\quad\quad]{
        \includegraphics[width=.552\linewidth]{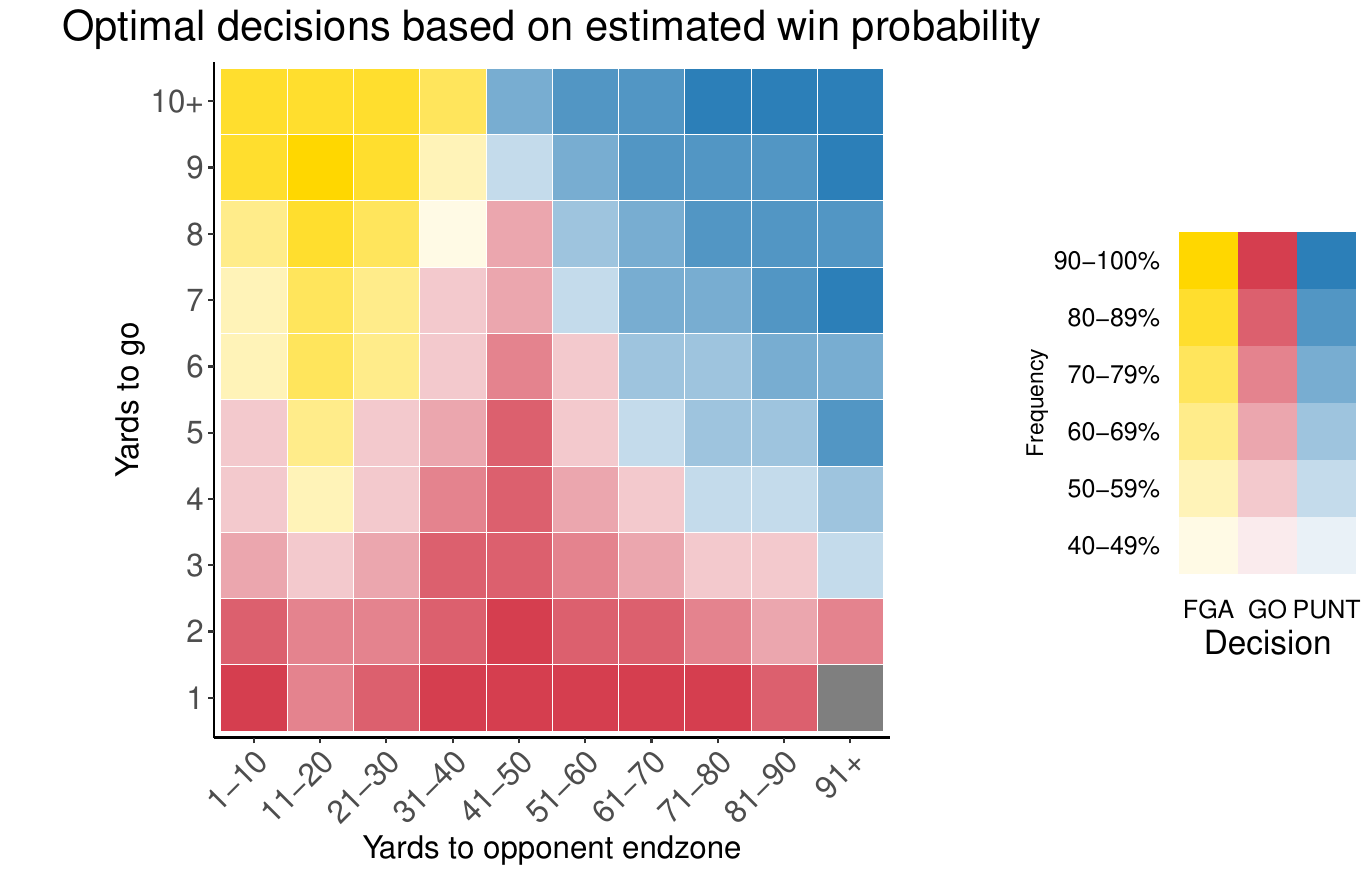}
        \label{fig:optimal_dec}
    }
\caption{\textbf{(a)} Most frequent fourth down decisions over the 2014-2022 NFL seasons with respect to yardline ($x$-axis) and yards to first down ($y$-axis).  Decisions are represented by fill color: yellow denotes field goal attempt (FGA), red denotes go for it (GO), and blue denotes punt (PUNT). The amount of color saturation corresponds to the frequency of the decision---darker colors represent higher frequencies.  \textbf{(b)} Most frequent fourth down decision prescriptions based on \cite{nfl4th} with respect to yardline and yards to first down.  Here, color saturation corresponds to the frequency with which the relative majority decision is estimated to be optimal.  This figure is inspired by similar figures in \cite{burke2014} and \cite{baldwin4thbot}.} 
\label{fig:observed_vs_optimal}
\end{figure}

Prior research on the fourth down decision demonstrates that empirical fourth down behavior is inconsistent with theoretically optimal decisions \citep{carter1978note, romer2006firms, burke2014, owens2018decision, yam2018fourthdown, baldwin4thbot}.  Figure \ref{fig:observed_vs_optimal} illustrates the gap between league-average fourth down behavior and an estimated optimal decision map. Panel (a) shows what coaches in the NFL did most often on fourth down over the 2014 to 2022  seasons, as a function of the yardline (in 10-yard intervals) and yards to go until a first down.  Panel (b) shows the optimal actions with respect to win probability estimates from a statistical model \citep{nfl4th}.\footnote{This win probability model includes additional features that influence the estimated optimal decision in a given state besides the yardline and yards to first down.    In order to illustrate the model's prescriptions in two dimensions, we show the decision that is most frequently optimal based on these two features, aggregating over all other covariates in the model (e.g. score differential, time remaining, timeouts remaining, etc.).  These win probabilities can be accessed via the R package \texttt{nfl4th} \citep{nfl4th}.}   

While prior research on the fourth down decision has demonstrated that coaches are not risk-neutral decision makers on fourth down, to our knowledge no one has attempted to estimate a risk profile that best explains their behavior.  Our goal is to estimate the implicit risk sensitivity of coaches' on fourth downs based on their observed decisions.  To do this, we assume that coaches optimize with respect to an unknown \textit{quantile} of the future game value distribution.  Mathematically, letting $C(X,a)$ denote the random variable for future game value conditional on taking action $a$ in game situation $X$, instead of choosing an action that maximizes $\mathbb{E}[C(X,a)]$, the coach maximizes $\mathbb{Q}_{\tau}[C(X,a)]$, the $\tau$-quantile of $C(X,a)$.

\begin{figure}
\centering
    \subfloat[]{
        \includegraphics[width=.47\linewidth]{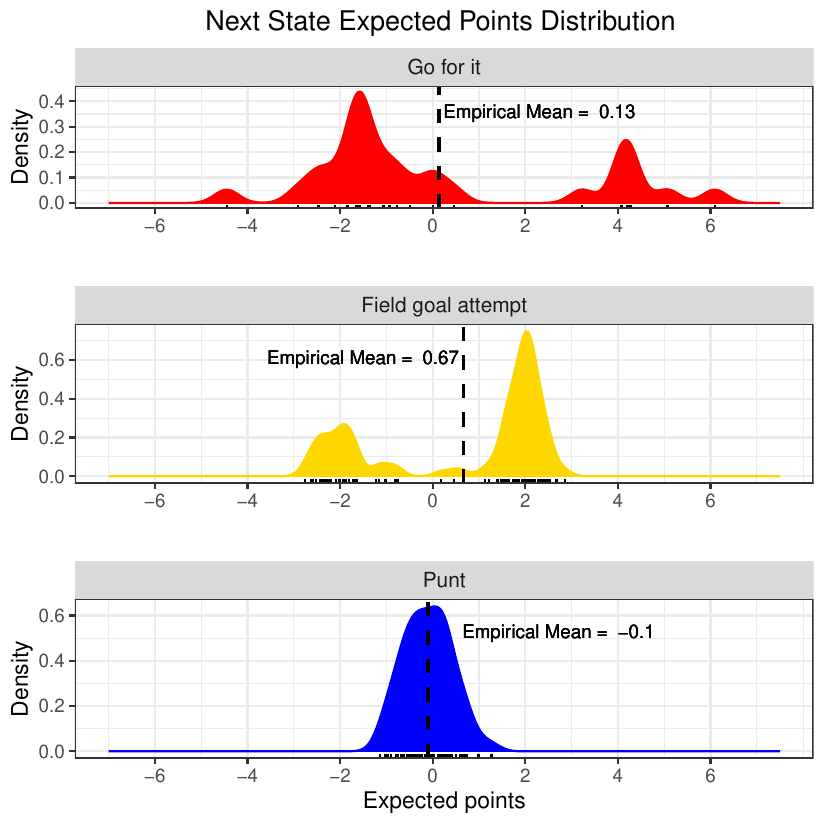}
        \label{fig:nsavd}
    }
    \hspace{1em}%
    \subfloat[]{
        \includegraphics[width=.47\linewidth]{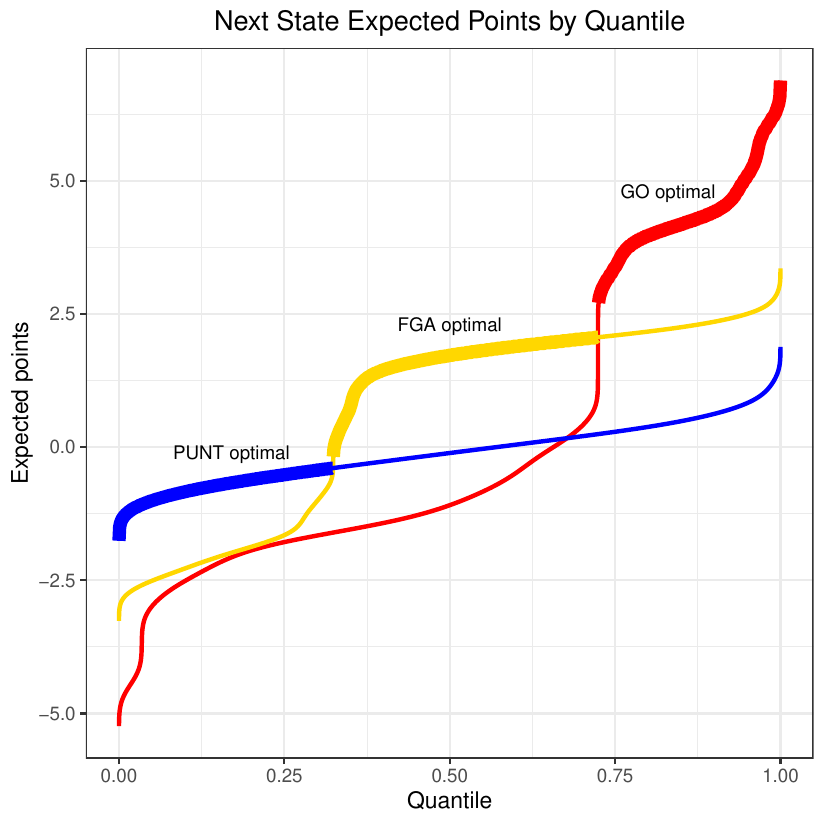}
        \label{fig:nsav_qf}
    }
\caption{\textbf{(a)} KDEs of the immediate reward plus next state expected points distribution associated with each fourth down decision (from the perspective of the team with possession), conditional on the game situation being fourth down at the 38 yardline, with 4 yards to first down.  The expected point estimates for the observed outcomes are shown by rugs in each plot. The dashed black lines denote the empirical mean expected points for each action's observed outcomes. \textbf{(b)} The quantile functions for the KDE of each action's next state expected points distribution (GO-red, FGA-yellow, PUNT-blue).  Each curve is bolded over the quantiles for which the action is optimal.}
\label{fig:inverse_illustration}
\end{figure}

Figure \ref{fig:inverse_illustration} shows how the optimal decision on fourth down can substantially change based on the quantile being optimized.  Panel (a) shows kernel density estimates (KDE) of the next state expected points distributions (plus any points incurred on the initial play) associated with each fourth down decision, conditional on the game situation being 4th down at the 38 yardline, with 4 yards to first down.\footnote{These expected points distributions are based on the model output from \cite{nflfastR}.}  In this situation, going for it, attempting a field goal, and punting are all viable options.  Panel (b) shows the quantile functions associated with the KDEs from panel (a).  The bolded portion of each curve highlights the quantiles over which that decision is optimal. For low quantiles (i.e., more conservative risk preferences), punting is optimal.  As the quantile rises (i.e., risk-tolerance increases), the optimal decision changes to field goal attempt, then to going for it.  
 
At its core, our methodology amounts to inverse optimization.  In inverse optimization (IO), prescribing behavior is \textit{not} the goal.  Instead, observed decisions are assumed to be optimal while the optimization model (or parameters thereof) is latent.    The objective is to estimate an optimization model such that the observed actions are optimal (or minimally suboptimal) with respect to this model \citep{chan2023inverse}.  

In this paper, we model the fourth down decision (and the subsequent game states) as a Markov decision process (MDP), where the coach's observed decisions are assumed to be optimal with respect to an unknown quantile of future value.  We then construct a framework to estimate these quantiles via inverse optimization.  This provides a way to explain coaches' behavior on fourth down in quantifiable terms that directly map to risk.  More broadly, to the best of our knowledge, we are the first to consider inverse optimization of quantile MDPs.

\subsection{Data and Code}

We use publicly available NFL play-by-play data over nine seasons (2014 through 2022).  These data are accessible via the R package \texttt{nflfastR} \citep{nflfastR}. They include detailed contextual information at the play level, such as down, yards to go, yardline, play description, play type, time remaining and score differential. The dataset also includes derived features (e.g., length of a drive) and model outputs (e.g., estimated win probability) associated with each play. 

R code to reproduce the results and figures in this paper is hosted publicly at the first author’s GitHub page: \url{ https://github.com/nsandholtz/fourth_down_risk}.

\section{Related work} \label{sec:lit_review}

\subsection{Risk-sensitive decisions in football}

Existing literature on the fourth down decision demonstrates that NFL coaches generally do not behave optimally with respect to either risk-neutrality or win probability \citep{romer2006firms, burke2014, owens2018decision, baldwin4thbot}.  \cite{yam2018fourthdown} builds on these findings, aiming to estimate the actual value that was lost (in terms of wins) by the league's overly conservative fourth down behavior.  They estimate that suboptimal fourth down behavior costed teams approximately 0.4 wins per year on average.  Recent work by \cite{roach2024updating} finds that within-game prior fourth down successes and failures influence coaches subsequent behavior in the game.  They find that coaches are, on average, more sensitive to failures than successes, which leads to suboptimal fourth down decisions later in the game. 

While recent studies have confirmed that coaches are indeed overly conservative when making fourth down decisions, they reveal a more nuanced narrative.  \cite{grafstein2023correcting} shows that estimated optimal fourth down policies can be corrupted by selection bias in the raw data, leading to overly aggressive policies.  \cite{lopez2020bigger} additionally finds that estimated policies tend to be overly aggressive due to confounding that arises from rounding the exact amount of yardage needed to obtain a first down to integer values.\footnote{For example, the play states "4th down and 2 inches to go" and "4th down and 71 inches to go" are both mapped to "4th down and 1 yard to go” in the raw play-by-play data, but both coach behavior and the corresponding average success rates vary significantly between these two situations.}  In a separate vein, \cite{brill2023analytics} finds that the uncertainty is often understated in win probability-based fourth down policy estimates.  

Regarding a different risk-sensitive football decision, \cite{urschel2011nfl} explore the risk preferences of NFL coaches on kickoff decisions using the framework of expected utility theory and prospect theory \citep{von2007theory, kahneman2013prospect}.  Our work is similar in that we propose a methodology to yield specific estimates of the magnitude of a team's risk-sensitivity based on their observed behavior.  However, our decision model is parameterized by the risk measure itself rather than by a \textit{utility function}, which is a transformation of the expected reward values.  While utility functions are a useful tool by which to contextualize risk, we feel that it is valuable to approach this problem by directly inferring parameters of the risk measure.  

The aforementioned work of \cite{grafstein2023correcting} uses a Heckman selection model to correct for preferential bias in the raw play-by-play data when estimating fourth down transition probabilities.  They include a coaching random effect in their model, thereby providing estimates of coaching ``preferences" on fourth down independent of the game situation and team ability.  While these random effects do not objectively measure risk, they do provide a way to compare the relative propensity of a coach's tendency to go for it on fourth down.   

Other risk-sensitive football decisions have been studied besides the fourth down decision.  Multiple studies find that coaches are not risk-neutral with respect to the decision of whether to pass or run the ball, generally concluding that teams under-utilize passing, which is the riskier of the two decisions \citep{alamar2006passing, kovash2009professionals, alamar2010measuring, critchfield2015matching}.  Others have found evidence of risk-sensitivity in NFL draft decisions \citep{massey2013loser}.

\subsubsection{Learning risk preferences through inverse optimization}

While there is a robust literature on decision processes with risk sensitive objectives \citep{jaquette1973markov, sobel1982variance, white1988mean, bellemare2017distributional, gilbert2017optimizing, li2017quantile}, this body of work primarily studies forward problems rather than inverse problems.  In most inverse decision problems, the objective function is either deterministic, as in inverse optimization, or else the risk measure is assumed to be the expectation, as in inverse reinforcement learning and revealed preference methods \citep{ng2000algorithms, varian2006revealed}.  

There are relatively few works that utilize inverse optimization frameworks to directly learn parameters of a risk measure.   \cite{yu2020learning} proposes an IO method for measuring risk preferences from financial portfolios by augmenting the objective by the covariance matrix of asset level portfolio returns.   \cite{di2019practical} introduces `risk-shaping', which applies inverse reinforcement learning (IRL) to infer the risk measure over the cumulative reward of an MDP that yields a given policy as optimal (rather than inferring the reward function, as is the standard in IRL).   \cite{li2021inverse} uses inverse optimization to learn a convex function representative of an investor's risk preferences.  \cite{sandholtz2023inverse} proposes methods to estimate how humans synthesize risk when searching for the optimum of a latent objective function, based on observed search paths from an experiment.   

There is also a significant body of literature devoted to learning risk preferences that do not explicitly leverage IO. However, as IO is central to the methodology we propose, we do not focus on connections to this body of work.


\section{Decision model} \label{sec:decision_model} 

We begin by formulating the fourth down decision as a MDP.  This development is prerequisite to the formulation of the inverse problem, which is considered in Section \ref{sec:inverse_problem}.  

\subsection{MDP components}

\begin{figure}
\begin{center}
\includegraphics[width=1\textwidth]{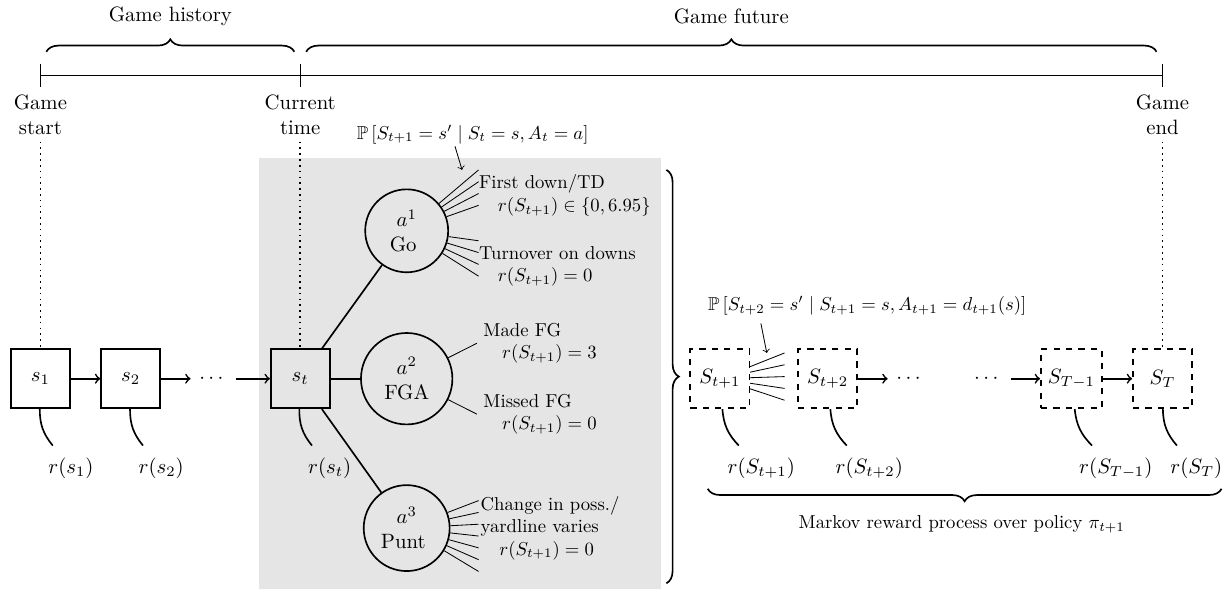}
\end{center}
\caption{Graphical model of the fourth down decision in the context of the game timeline.   Solid square nodes represent observed game states, circular nodes denote the possible actions that can be taken on fourth down, and curved lines emerging from state nodes represent state-dependent rewards.  Dashed square nodes represent future (unobserved) game states, with corresponding rewards emerging from these future states.  The period in which the decision takes place is highlighted by a gray box, and the most common outcomes for each action are listed (e.g., first down or touchdown for action $a^1= \text{Go}$).  The lines connecting actions to the listed outcomes represent state/action-dependent transition probabilities.  The lines connecting future state $S_{t+1}$ to $S_{t+2}$ likewise represent state/action-dependent transition probabilities, however, here we represent the action via decision rule $d_{t+1}$.  Given policy $\pi_{t+1}$, future states and rewards follow a Markov reward process.} \label{fig:decision_tree}
\end{figure}

Figure \ref{fig:decision_tree} shows a graphical depiction of our model of the fourth down decision in the context of the game timeline.  
Assuming the current decision epoch is $t$, $d_{t+1}$ is a \textit{decision rule} or a mapping from states to actions at time $t+1$, and $\pi_{t+1}$ is a \textit{policy}, or a collection of decision rules governing the remaining actions in the game: 
\begin{align}
    d_{t+1}: \mathcal{S}(t+1) \rightarrow \mathcal{A}(t+1) \nonumber \\
    \pi_{t+1} = \{d_{t+1}, d_{t+2}, \ldots, d_T\}, \nonumber
\end{align}
where $\mathcal{S}(t+1)$ and $\mathcal{A}(t+1)$ are the sets of feasible states and actions respectively, in epoch $t+1$.  When policy $\pi_{t+1}$ is fixed, the process becomes a one-period MDP where game states after play index $t+1$ evolve according to a Markov reward process.  These future states are shown in upper case in Figure \ref{fig:decision_tree} (e.g., $S_{t+1}$) as they are random variables, while the states that have already been observed are denoted using lower case (e.g., $s_{1}$).  The index $T$ denotes the final play of the game and is itself a random variable.  In principle, the fourth down decision can be modeled as a multi-period Markov decision process, enabling the possibility of optimizing future decisions. While this represents a compelling direction of future research, this is not the focus of our paper.  The literature universally adopts the one-period perspective for tractability and we do the same. 

\subsubsection{States}

We define the state space, $\mathcal{S}$, to be the union of two mutually exclusive sets: $\mathcal{S}^{\text{score}}$, the set of scoring states, and $\mathcal{S}^{\text{play}}$, the set of game states at the start of a play. A scoring state is a tuple consisting of the team that scored (denoted generically as either team A or B) and the type of score (touchdown, field goal or safety):
\begin{align}
    \mathcal{S}^{\text{score}} = \{\text{A}, \text{B}\} \times \{\text{TD}, \text{FG}, \text{SAF}\}. \nonumber
\end{align} 
A play state is a tuple consisting of the team with possession, down, yardline (binned into groups of 10 yards), and yards to go until first down:
\begin{align}
    \mathcal{S}^{\text{play}} = \{\text{A}, \text{B}\} \times \{1,2,3,4\} \times \{(1-10),\ldots,(91-99)\} \times \{1,2, \ldots, 9, 10+\}. \nonumber
\end{align}

The state variable $s$ will be used to refer to an arbitrary game state in $\mathcal{S}$. Let $\mathcal{S}^4 \subseteq \mathcal{S}^\textrm{play}$ denote the set of fourth down play states. Going forward, elements of $\mathcal{S}^4$ will be denoted $\sigma$.

\subsubsection{Actions}

Given that we treat future actions as deterministic according to fixed policy $\pi_{t+1}$, the only action set we are concerned with is the set of fourth down actions.  Hence $\mathcal{A} = \{\text{GO}, \text{FGA}, \text{PUNT}\}$, denoting go for it, field goal attempt, and punt, respectively.  The set of realistic options for a given fourth down situation is typically either $\{\text{GO}, \text{FGA}\}$ or $\{\text{GO}, \text{PUNT}\}$, depending on the field position.  We use $\mathcal{A}$ for simplicity.  
 
 \subsubsection{Transition probabilities} \label{sec:trans_prob_def}
 
We require notation for the probability of transitioning to an arbitrary state in $\mathcal{S}$ conditional on a given fourth down state and action. Let 
\begin{align} \label{eq:tp_state_action}
p(s' | \sigma, a) &:=\mathbb{P}\left[S_{t+1}=s^{\prime} \mid S_{t}=\sigma, A_{t}=a\right], 
\end{align}
where $s' \in \mathcal{S}, \sigma \in \mathcal{S}^4$, and $a \in \mathcal{A}$.  We also need notation for the transition probabilities governing the remainder of the game under fixed policy $\pi_{t+1}$.  Let 
\begin{align} \label{eq:tp_state_policy}
    p(s' | s, \pi_{t+1}) &:=\mathbb{P}\left[S_{t+i+1}=s^{\prime} \mid S_{t+i}=s, A_{t+i} = d_{t+i}(s)\right],
\end{align}
where $i \in 1,2, \ldots, {T-(t+1)}$ and all other terms are as defined previously.  The estimation of these probabilities is discussed in Section \ref{sec:inference}.

\subsubsection{Reward function}

We define the reward function from the perspective of team A:
\begin{align}
r(s) =  \begin{cases} 
      6.95, & s = (\text{A, TD}) \in \mathcal{S}^{\text{score}}, \\
      3, & s = (\text{A, FG}) \in \mathcal{S}^{\text{score}}, \\
      -2, & s = (\text{A, SAF}) \in \mathcal{S}^{\text{score}}. \\
  \end{cases} \label{eq:reward_function}
\end{align}
If team B scores, the reward values are the negative of the values in equation \eqref{eq:reward_function}.\footnote{Since our state-space includes states that exclusively represent non-zero rewards (i.e., scores), we can define the reward function as depending on the state variable alone.  This is unlike a typical MDP, in which the reward function depends on both states and actions.}  Note that we do not explicitly model the extra point or two-point conversion play after a touchdown.  As in \cite{romer2006firms} and \cite{chan2021points}, we instead define the touchdown reward as six points plus the league-average point value over all one- and two-point conversion attempts, which in our data set is 6.95 points.  

\subsection{The forward problem}

Given the MDP defined above, the fourth down decision becomes an optimization problem of the form
\begin{equation}
    \max_{a \in \mathcal{A}} q^{\pi_{t+1}}_t(\sigma, a), \label{eq:basic_forward_problem}
\end{equation}
where the objective function $q^{\pi_{t+1}}_t$ represents the long-run value (e.g., expected points, win probability, etc.) associated with taking action $a$ in fourth down state $\sigma$ at time $t$ and following $\pi_{t+1}$ thereafter.   In our analysis, we fix $\pi_{t+1}$ at the \emph{league-average policy} $\bar\pi$, assumed to be stationary. 

We recognize that in many instances our MDP formulation may not match the actual in-game optimization problem faced by a coach.  For example, to maintain computational feasibility and manage data sparsity issues, we excluded variables such as score differential, time remaining, and timeouts remaining from the state space, which often factor into coaches' fourth down decisions.  To mitigate this limitation, we stratify our analyses in Section \ref{sec:results} by win probability, which acts as a univariate proxy for these omitted variables.

Various functional forms have been proposed for $q^{\bar\pi}$ in the literature, including next score expected points \citep{carter1971operations, goldner2012markov, yurko2019nflwar}, extended or infinite horizon expected points \citep{winston1983technical, romer2006firms, chan2021points}, and win probability \citep{burke2014, yurko2019nflwar, baldwin4thbot}.  While there are differences in the prescribed decisions depending on which of these characterizations is used, each one results in a collection of prescriptions that are quite dissimilar from the coaches' observed behavior.  In the inverse problem, we aim to infer $q^{\bar\pi}$ such that the resulting prescriptions closely match the coaches' observed behavior.


\section{The inverse problem} \label{sec:inverse_problem} 

Formally, given a vector of $N$ decisions $\mathbf{a} := (a_1, \ldots, a_N)$ that were observed in $N$ corresponding fourth down states $\boldsymbol \sigma := (\sigma_1, \ldots, \sigma_N)$, we formulate the inverse problem as identifying a function $q^{\bar\pi} \in \mathcal{Q}^{\bar\pi}$ that minimizes a given loss function $\ell(\cdot,\cdot)$:
\begin{equation}\label{eq:approx_inv}
    \min_{q^{\bar\pi} \in \mathcal{Q}^{\bar\pi}} \ell(\mathbf{a}, \mathbf{a}^*(\boldsymbol \sigma, q^{\bar\pi})). 
\end{equation} 
The function class $\mathcal{Q}^{\bar\pi}$ contains candidate objective functions that summarize the long-run value of taking a particular action in a particular fourth down state and following $\bar\pi$ thereafter. Let vector $\mathbf{a}^*(\boldsymbol \sigma, q^{\bar\pi}) := (a_1^*(\sigma_1, q^{\bar\pi}), \ldots, a_N^*(\sigma_N,q^{\bar\pi}))$ be the vector of optimal decisions with respect to $q^{\bar\pi}$ in states $\boldsymbol \sigma$, with components defined as
\begin{equation}\label{eq:astar}
a_j^*(\sigma_j, q^{\bar\pi}) \in \argmax_{a \in \mathcal{A}} q^{\bar\pi}(\sigma_j,a), \quad j = 1, \ldots, N. 
\end{equation}
In the next two subsections, we define the class of candidate objective functions $\mathcal{Q}^{\bar\pi}$ and the loss function $\ell(\cdot,\cdot)$. 

\subsection{Constructing \texorpdfstring{$\mathcal{Q}^{\bar\pi}$}{Lg}} \label{sec:cand_obj_functions}

To construct $\mathcal{Q}^{\bar\pi}$ we will leverage the quantity referred to as the \textit{next-state value} \citep{bellemare2017distributional}, which is the random variable defined by
\begin{equation} \label{eq:quant_exp2}
    V_t^{\bar\pi}(\sigma,a) := r(S_{t+1}(\sigma,a)) + \mathbb{E}_{\bar\pi}\left[\sum \limits_{n=t+2}^T r(S_n) \,\middle\vert\, S_{t+1}(\sigma,a)\right], 
\end{equation}
where $S_{t+1}(\sigma,a)$ denotes the random variable for the next state conditional on taking action $a$ in current fourth down state $\sigma$.  The expectation in \eqref{eq:quant_exp2} is typically called the \textit{value function} at time $t+1$ for policy $\bar\pi$.  To simplify subsequent expressions, we will use the following notation for the value function:
\begin{equation}
v_{t}^{\bar\pi}(s) := \mathbb{E}_{\bar\pi}\left[\sum \limits_{n=t+1}^T r(S_n) \,\middle\vert\, S_{t} = s\right]. \label{eq:value_func}
\end{equation}
In our context, the next-state value denotes the sum of the reward from the next state visited after fourth down state $\sigma$ when action $a$ is chosen, plus the expected point differential over the remainder of the game under policy $\bar\pi$. Given our assumption that $\bar\pi$ is stationary, we will omit the subscript from $V_t^{\bar\pi}$ and $v_{t+1}^{\bar\pi}$ going forward. 

Note that $V^{\bar\pi}(\sigma,a)$ has a discrete and finite support since there are only a finite number of states to which the play can immediately transition. Specifically, its support consists of $r(s') + v^{\bar\pi}(s')$ for all $s' \in \mathcal{S}$ and the corresponding probability mass for each value in the support is $p(s' | \sigma, a)$:
 \begin{equation}
{V}^{\bar\pi}(\sigma, a)=
    \begin{cases}
        r(s_1) + {v}^{\bar\pi}(s_1) & \text{with prob.  } {p}(s_1 | \sigma,a),\\
        ~~~~~~~~\vdots & ~~~~~~~~~ \vdots\\ 
        r(s_{|\mathcal{S}|}) + {v}^{\bar\pi}(s_{|\mathcal{S}|}) & \text{with prob.  } {p}(s_{|\mathcal{S}|} | \sigma,a). \label{eq:nsav_distribution}
    \end{cases}
\end{equation}

In order to parameterize the inference space with respect to the risk measure, instead of taking the expectation of the next-state value, we employ the quantile function. We could use different risk measures to parameterize risk, such as spectral risk measures or variations on value at risk, but we use the quantile function because it is simple to interpret, flexible enough for our application, and easily maps to risk. Let $q^{\bar\pi}_\tau(\sigma,a)$ denote the $\tau$-quantile of $V^{\bar\pi}(\sigma,a)$:  
\begin{align}
    q^{\bar\pi}_\tau(\sigma,a) &= \mathbb{Q}_\tau\left[V^{\bar\pi}(\sigma,a)\right] \nonumber \\ 
                               &= \inf \{x \in \mathbb{R}: \tau \leq F_{V^{\bar\pi}}(x| \sigma,a)\}.\label{eq:quant_exp1}
\end{align} 
where $F_{V^{\bar\pi}}(x| \sigma,a)$ is the distribution function of random variable $V^{\bar\pi}(\sigma,a)$. 
Finally, we define the class of candidate objectives $\mathcal{Q}^{\bar\pi}$ by indexing \eqref{eq:quant_exp1} over $\tau \in [0,1]$:
\begin{equation}
\mathcal{Q}^{\bar\pi} = \left\{q^{\bar\pi}_\tau(\cdot,\cdot) \,\middle\vert\, \sigma \in \mathcal{S}^4, a \in \mathcal{A}, \tau \in [0,1] \right\}.\label{eq:inference_space}
\end{equation}

Characterizing future value via the next-state value distribution may seem like an odd choice compared to the distribution of the return  (i.e., the sum in \eqref{eq:value_func}).  However, using the distribution of the return would inherently assume that coaches operate according to the same quantile in all future game situations, which is not plausible.  Since we know that coaches' risk tolerances are uniquely different on fourth down decisions than in other game situations, we believe that \eqref{eq:quant_exp1} represents a more realistic objective.  In other words, it does not make sense to us to consider the objective 
\begin{equation}
\mathbb{Q}_\tau\left[\sum \limits_{n=t+1}^T r(S_n) \,\middle\vert\, S_t = \sigma, A_t = a\right],
\end{equation}
since this would assume coaches operate with respect to the same quantile objective for not only the immediate fourth down decision, but all future decisions. We therefore posit that 
\begin{equation}
\mathbb{Q}_\tau\left[ r(S_{t+1}(\sigma,a)) + \mathbb{E}_{\bar\pi}\Bigg(\sum \limits_{n=t+2}^T r(S_n)\Bigg) \,\middle\vert\, S_t = \sigma, A_t = a\right],
\end{equation}
represents a more realistic objective.  

Before proceeding, we pause to acknowledge that \eqref{eq:inference_space} almost certainly cannot not perfectly describe any coach's objective function.  Indeed, a coach's "true" objective function may well be some combination of win-probability, point-efficiency, social conformity, prior experience, and any number of other factors.  However, we find that the fitted quantile objective functions generally explain coaches' behavior well (see Section \ref{sec:results}) and furthermore, the estimated quantiles convey meaningful information in terms of risk, which is a significant factor in the fourth down decision. 

\subsection{Loss function}

We use average Hamming distance to measure loss \citep{hamming1950error}. Formally, 
\begin{equation}
    \ell(\mathbf{a}, \mathbf{a}^*(\boldsymbol \sigma, q^{\bar\pi}_\tau)) = \frac{1}{N}\sum_{j=1}^N \mathbbm{1}\big(a_j \neq a_j^*(\sigma_j, q^{\bar\pi}_\tau)\big)
\end{equation}
where $\mathbbm{1}(\cdot)$ is the indicator function and $a_j^*(\sigma_j, q^{\bar\pi}_\tau)$ is defined as in \eqref{eq:astar}. Under this loss function, an observed action $a_j$ incurs zero loss if it is optimal with respect to the quantile-based objective function associated with the corresponding fourth down state $\sigma_j$. 

\subsection{Inverse model}

With our specific choices of $\mathcal{Q}^{\bar\pi}$ and $\ell(\cdot,\cdot)$, the general inverse problem in \eqref{eq:approx_inv} becomes 
\begin{equation}\label{eq:singletau}
    \min_{\tau \in [0,1]} \frac{1}{N} \sum_{j=1}^N  \mathbbm{1}\big(a_j \neq a_j^*(\sigma_j, q^{\bar\pi}_\tau)\big)
\end{equation}
where $q_\tau^{\bar\pi}$ and $\mathcal{Q}^{\bar\pi}$ are as defined in \eqref{eq:quant_exp1} and \eqref{eq:inference_space} respectively. This problem chooses a single $\tau$ (i.e., a single function $q_\tau^{\bar\pi}$) that maximizes the ``fitness'' between the observed actions $a_j, j = 1, \ldots, N$ and $\tau$-optimal actions $a_j^*(\sigma_j, q_\tau^{\bar\pi}), j = 1, \ldots, N$ in corresponding fourth down states $\sigma_j, j = 1, \ldots, N$.  

In practice, it may be that a decision maker has different risk tolerances for different subsets of the state space \citep{kahneman2013prospect}.  This motivates an extension to formulation~\eqref{eq:singletau}, where we estimate multiple quantiles. Suppose there are $L \leq |\mathcal{S}^4|$ objective functions associated with quantiles $\tau_1, \ldots, \tau_L \in [0,1]^L$. Let $\mathcal{P}_1, \ldots, \mathcal{P}_L$ be a partition of the fourth down states (i.e., $\bigcup_{\ell = 1}^L \mathcal{P}_\ell = \mathcal{S}^4$ and $\bigcap_{\ell = 1}^L \mathcal{P}_\ell = \emptyset$). All states in the subset $\mathcal{P}_\ell$ will be associated with the same quantile $\tau_\ell$. Then, the inverse optimization problem can be written as

\begin{equation}\label{eq:multitau2}
    \min_{(\tau_1, \mathcal{P}_1), \ldots, (\tau_L, \mathcal{P}_L) } \frac{1}{N} \sum_{\ell=1}^L\sum_{\{j : \sigma_j \in \mathcal{P}_\ell\}} \mathbbm{1}\big(a_j \neq a_j^*(\sigma_j, q^{\bar\pi}_{\tau_\ell})\big),
\end{equation}
where $\tau_\ell \in [0,1]$ and $\bigcup_{\ell = 1}^L \mathcal{P}_\ell$ is a partition of $\mathcal{S}^4$.


\section{Inference} \label{sec:inference}

Our inverse optimization methodology requires estimates for the transition probabilities, value function, and quantiles of the next-state value distribution.  We also require a partitioning of the fourth down state space.  In this section we discuss how we construct this partitioning and estimate the aforementioned quantities.  

\subsection{Transition probabilities}

We estimate the state-action transition probabilities in \eqref{eq:tp_state_action} by the empirical proportions:
\begin{equation}
    \hat{p}(s' | \sigma,a) = \frac{\sum_{g \in \mathcal{G}} \sum_{t \in \mathcal{T}_g} \mathbbm{1} \Big(s_t = \sigma, a_t = a, s_{t+1} = s'\Big)}{\sum_{s' \in \mathcal{S}} \sum_{g \in \mathcal{G}} \sum_{t \in \mathcal{T}_g} \mathbbm{1} \Big(s_t = \sigma, a_t = a, s_{t+1} = s'\Big)}, \label{eq:lambda_hat}
\end{equation}
where $\mathcal{G}$ is the set of games in our dataset and $\mathcal{T}_g$ is the set of all play indices in game $g$.  We likewise estimate \eqref{eq:tp_state_policy}, the state transition probabilities given policy $\bar\pi$,  empirically:
\begin{equation}
    \hat{p}(s' | s, \bar\pi) = \frac{\sum_{g \in \mathcal{G}} \sum_{t \in \mathcal{T}_g} \mathbbm{1} \Big(s_t = s, s_{t+1} = s'\Big)}{\sum_{s' \in \mathcal{S}} \sum_{g \in \mathcal{G}} \sum_{t \in \mathcal{T}_g} \mathbbm{1} \Big(s_t = s, s_{t+1} = s'\Big)}. \label{eq:theta_hat}
\end{equation}

Note that we could compute \eqref{eq:lambda_hat} and \eqref{eq:theta_hat} based on additional criteria in the indicator functions.  For example, we could restrict the analysis to plays from a specific season, or to plays within a fixed interval of estimated win probability.  We could then solve the inverse problem with these additional restrictions to study risk preferences in specific situations of interest.  Indeed, this is what we do in Section \ref{sec:results}---we explore how risk preferences vary by win probability, season, coach and team.   However, for simplicity we omit these additional features in the notation.   

\subsection{Value function}

The value function, defined in \eqref{eq:value_func}, may be regarded as the relative expected point advantage (or disadvantage) of starting in state $s$ over the remaining $T-t$ plays in the game.  Since the steady state is reached quickly in a recurrent Markov chain in football \citep{chan2021points}, this can be closely approximated by taking the expected return over the infinite horizon:
\begin{equation}
v^{\bar\pi}(s) \approx \mathbb{E}_{\bar\pi}\left[\sum \limits_{n=t+1}^\infty r(S_n) \,\middle\vert\, S_{t} = s\right]. \label{eq:value_func_approx}
\end{equation}
As shown in \cite{chan2021points}, by treating the game as between two identical league-average teams, \eqref{eq:value_func_approx} can be analytically derived given the transition dynamics and reward function.  Following their methodology, we use the empirical transition probabilities in \eqref{eq:theta_hat} and the reward function defined in \eqref{eq:reward_function} to obtain an estimate of the value function.\footnote{Supplement A provides a brief explanation of analytical derivation of \eqref{eq:value_func_approx} and a figure with the corresponding estimated value function for our application.}

 \subsection{Quantiles of next-state value} \label{sec:next_state_quantiles}
  
 With estimates of $p(s' | \sigma,a)$ and $v^{\bar\pi}(s')$, we can plug these into \eqref{eq:nsav_distribution} to approximate the next-state value distribution for any state-action pair $(\sigma, a) \in \{\mathcal{S}^4 \times \mathcal{A}\}$.  Then, to construct $\mathcal{Q}^{\bar\pi}$ as given in \eqref{eq:inference_space}, we take quantiles over $\hat{V}^{\bar\pi}(\sigma,a)$.  For a given fourth down state and action, we estimate the $\tau$-quantile of $\hat{V}^{\bar\pi}(\sigma,a)$ as:
\begin{align}
    \hat{q}_{\tau}^{\bar\pi}(\sigma, a) = \inf \{x \in \mathbb{R}: \tau \leq \hat{F}_{V^{\bar\pi}}(x)\}, \label{eq:quantile_action_value_est}
\end{align}
where $\tau \in (0,1)$ and $\hat{F}_{V^{\bar\pi}}$ is the empirical distribution function of $\hat{V}^{\bar\pi}(\sigma, a)$.  

\begin{figure}
\centering
    \sidesubfloat[]{
        \includegraphics[width=.82\linewidth]{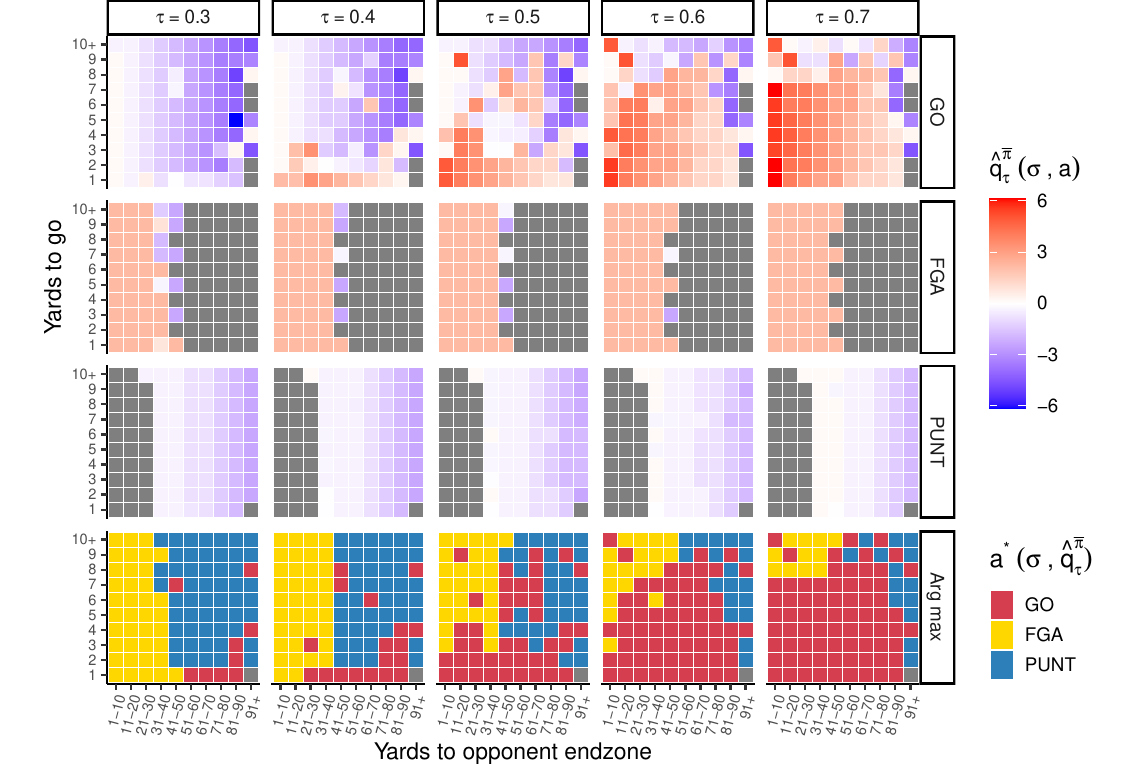}
        \label{fig:quantile_action_values_raw}
    }
    \vspace{1em}%
    \vspace{1em}%
    
    \sidesubfloat[]{
        \includegraphics[width=.82\linewidth]{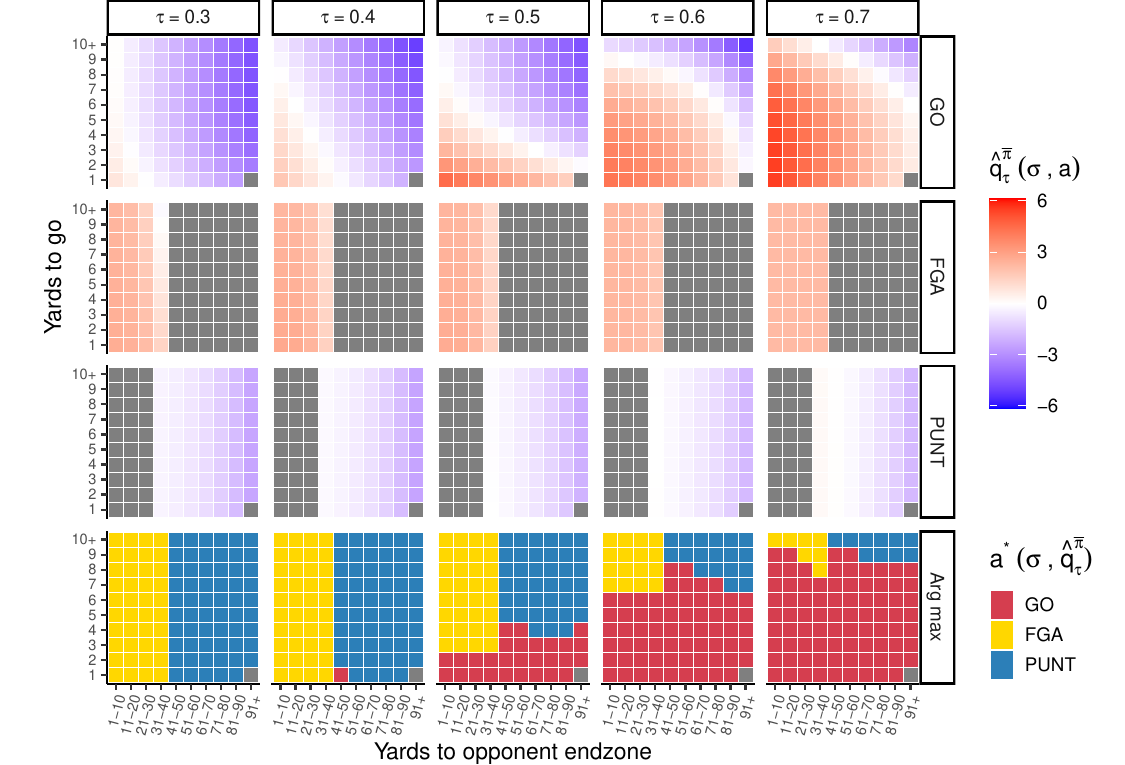}
        \label{fig:quantile_action_values_smooth}
    }
\caption{\textbf{(a)} Top three rows: Estimates of ${q}_{\tau}^{\bar\pi}(\sigma, a)$ over all $(\sigma, a)$ combinations for $\tau \in \{.3, .4, .5, .6, .7\}$.  Within plots, the horizontal and vertical axes correspond to yards to opponent end zone and yards until first down, respectively.  Across plots, the horizontal and vertical axes correspond to quantile and action, respectively. Bottom row: Decision rules, as given by \eqref{eq:astar}, for the $\hat{q}_{\tau}^{\bar\pi}(\sigma, a)$ values. \textbf{(b)} Identical to panel (a), but using the smoothed estimates of ${q}_{\tau}^{\bar\pi}(\sigma, a)$.}
\end{figure}

The top three rows of Figure \ref{fig:quantile_action_values_raw} show these estimates over all pairs $(\sigma, a) \in \{\mathcal{S}^4 \times \mathcal{A}\}$ for $\tau \in \{0.3, 0.4, 0.5, 0.6, 0.7\}$. The bottom row of Figure \ref{fig:quantile_action_values_raw} depicts the corresponding $\tau$-optimal decision maps (i.e., $a^*(\sigma,\hat{q}^{\bar\pi}_\tau)$ from \eqref{eq:astar} for all $\sigma \in \mathcal{S}^4$). Notice that the raw $\hat{q}_{\tau}^{\bar\pi}(\sigma, a)$ values in Figure \ref{fig:quantile_action_values_raw} vary drastically for the GO action (top row) in some regions of the state space.  This is because coaches rarely (or never) go for it in these situations, hence the corresponding empirical estimates of $p(s'|\sigma,a)$ are unstable.  The instability this causes propagates to the $\tau$-optimal decision maps, as shown by the existence of counter-intuitive go-for-it ``islands" in the bottom row of Figure \ref{fig:quantile_action_values_raw}.

To address these issues, we employ a bivariate monotonic smooth over $\hat{q}_{\tau}^{\bar\pi}(\sigma, a)$, which allows us to leverage monotonic relationships that we can reasonably assume over the state space. Specifically, for a given action $a$ and quantile $\tau$, we fit a shape-constrained additive model to each set $\left\{\hat{q}^{\bar\pi}_\tau(\sigma,a) \,\middle\vert\, \sigma \in \mathcal{S}^4\right\}$, where monotonic-decreasing smoothness constraints are jointly specified over the yardline and yards-to-go covariates via tensor product smoothing with bivariate penalized b-splines \citep{pya2015shape}.  We fit these models using the {\ttfamily scam} R package \citep{scam2020}.  We used $k=4$ knots as the basis dimensions for both the yardline and yards-to-go covariates.  Additionally, since the fourth down GO data is corrupted by selection bias \citep{grafstein2023correcting}, we augment the fourth down transition data for $a = $ GO with third down plays, as in \cite{romer2006firms}.  Specifically, for a given third down play, we determine what the realized value of the next state would have been if the play had taken place on fourth down, then supplement the fourth down data with these additional transitions.  

Figure \ref{fig:quantile_action_values_smooth} illustrates how the quantile estimates and the resulting decision maps change after applying these regularization steps.  The regularized decision maps are more intuitive; for each action, the set of states where the action is prescribed comprises a contiguous region.  As such there are no longer ``islands" in questionable states.  

\subsection{Partitioning \texorpdfstring{$\mathcal{S}^4$}{Lg}} \label{sec:partition_s4}

Next, we partition the fourth down state space, allowing the model to infer different risk preferences for different portions of the state space.  This requires choosing a value for $L$, the number of subsets in the partition, and determining the structure of the subsets.  Because of the small size of the action set, the risk preferences become severely underidentified for values of $L \geq 3$.  In fact, for all optimal partitions of size $L \geq 3$, the minimized loss as given by \eqref{eq:multitau2} is equal at the minimum value.  This is because once the partition size is equal to (or greater than) the size of the action space ($|\mathcal{A}| = 3$), a partitioning can be constructed where, within any given subset, each game state has the same majority observed action.  As long as this property is satisfied, \eqref{eq:multitau2} will be minimized, no matter how many additional subsets are created.  This phenomenon is illustrated in Figure \ref{fig:partition_fig} of Supplement A.  

In order to retain identifiability while still allowing for some degree of model complexity, we solve \eqref{eq:multitau2} with $L = 2$.  However, even for $L=2$, the number of unique partitions of $\mathcal{S}^4$ is $2^{99}$, which renders solving \eqref{eq:multitau2} by enumeration impossible.  Instead we restrict our search to specific contiguous groupings of the fourth down state space as described in Appendix \ref{sec:partition_appendix} in Supplement A.  Under this setting, the partitions created by splitting the state space by the 40 yardline vs. the 50 yardline are both approximately optimal.  For interpretability reasons, we choose the partition given by splitting the state space at the 50 yardline.  Hence in all results going forward, we use the partition $\{\mathcal{P}_1,\mathcal{P}_2\}$, where $\mathcal{P}_1$ comprises fourth down states for which the team has possession in opposing half of the field and $\mathcal{P}_2$ is the complement of this set with respect to $\mathcal{S}^4$ (i.e., fourth down states in the team's own half of the field).  

This partition is intuitive in that it separates the two kicking decisions; most kicking decisions with fewer than 50 yards to the endzone are field goal attempts, while most kicking decisions with more than 50 yards are punts.  Also, the 50 yardline approximately marks the point at which fourth down states go from having a negative expected value to positive expected value (see Figure \ref{fig:value_function} in Supplement A).  

\subsection{Inverse problem solution}

Given estimates for $q^{\bar\pi}_\tau(\sigma,a)$ and the above partition of $\mathcal{S}^4$, the inverse problem given by \eqref{eq:multitau2} can be solved, yielding estimates for $\boldsymbol{\tau} \in \{[0,1] \times [0,1]\}$ for any given collection of decisions $\mathbf{a} := (a_{1}, \ldots, a_{N})$ in corresponding fourth down states $\boldsymbol \sigma := (\sigma_{1}, \ldots, \sigma_{N})$.  Having fixed the partition over $\mathcal{S}^4$, we obtain the following estimator for $\boldsymbol{\tau}$:
\begin{equation}\label{eq:multitau_exact}
   \hat{\boldsymbol{\tau}} = \argmin_{\boldsymbol{\tau} \in \{[0,1] \times [0,1]\}}  \frac{1}{N}\sum_{\ell=1}^2 \sum_{\{j : \sigma_j \in \mathcal{P}_\ell\}} \mathbbm{1}\big(a_{j} \neq a_{j}^*(\sigma_{j}, \hat{q}^{\bar\pi}_{\tau_\ell})\big),
\end{equation}
where $\boldsymbol{\tau} := (\tau_1, \tau_2)$, and $\mathcal{P}_1$ and $\mathcal{P}_2$ are as defined in Section \ref{sec:partition_s4}.  

In our application, there are often multiple $\tau$'s which satisfy \eqref{eq:multitau_exact}.\footnote{See Figure \ref{fig:league_aggregate_solution} in Appendix \ref{sec:additional_figs} of Supplement A for an illustration.}  This is because the $\tau$-optimal policies---when considered independently for each field region---are constant for many subsets of [0,1].  We use the medians of the set of quantiles which satisfy \eqref{eq:multitau_exact} as point estimates, but this phenomenon highlights the importance of contextualizing point estimates with an appropriate quantification of the corresponding uncertainty. 

\subsection{Uncertainty quantification} \label{sec:uq}

To characterize the uncertainty around point estimates for $\tau_1$ and $\tau_2$, we leverage the bootstrap \citep{efron1992bootstrap}.  To maintain the dependencies that exist on plays both within and across drives, we take bootstrap samples at the game level, as in \cite{franks2016meta}.  Specifically, for each bootstrap sample, we resample the set of game indices (with replacement) then collect \textit{all} plays from the resulting set of games. 

For the $b$th bootstrap sample, we draw $n$ games from $\mathcal{G}$ with replacement (where $n = |\mathcal{G}|$), yielding  $\mathcal{G}_b = \{g^1_b,\ldots,g^n_b\}$.  We then compute the empirical transition probabilities via \eqref{eq:lambda_hat} and \eqref{eq:theta_hat} respectively, only we substitute $\mathcal{G}_b$ for $\mathcal{G}$.  Using these, we compute the bootstrapped versions of \eqref{eq:vector_value_func} and \eqref{eq:quantile_action_value_est}, ultimately providing $\mathbf{a}_b^{*}(\boldsymbol \sigma_b, \hat{q}^{\bar\pi}_{\tau,b})$ from \eqref{eq:astar} for all feasible $(\sigma, \tau)$ combinations.  Figure \ref{fig:full_quantile_policy} in Appendix \ref{sec:additional_figs} depicts how the $\tau$-optimal policies vary from bootstrap sample to bootstrap sample. 

\begin{figure}
    \centering
    \includegraphics[width=1\linewidth]{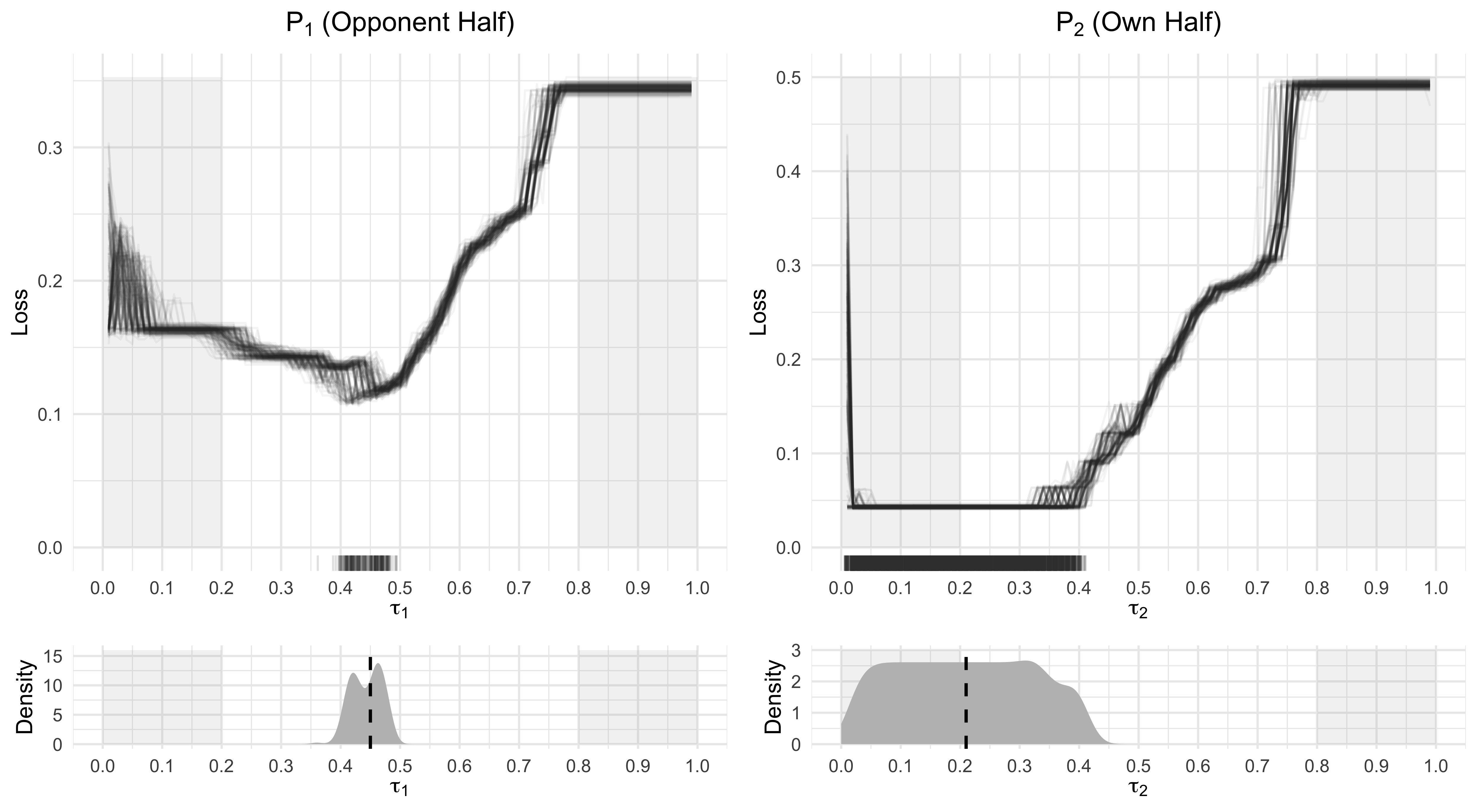}
    \caption{Top: loss curves for 200 bootstrap samples (black) for $\mathcal{P}_1$ (left) and $\mathcal{P}_2$ (right).  The rugs show the $\tau$ values that minimize each loss curve, or the collection of optimal $\tau$-sets.  Bottom: weighted KDEs of the optimal $\tau$-sets.  For each region, the dashed vertical line shows the weighted median over all 200 optimal $\tau$-sets.  The light gray shaded regions in each plot represent approximate areas where the model space is underidentified (i.e. $\tau < 0.2$ and $\tau > 0.8$).}
    \label{fig:bootstrap_loss_illustration}
\end{figure}

To incorporate this variability into the inference on $\boldsymbol \tau$, for each bootstrap sample $b$ we solve \eqref{eq:multitau_exact} with respect to $\mathbf{a}_b$ and $\boldsymbol \sigma_b$---the vectors of observed fourth down decisions and corresponding fourth down states in games $\mathcal{G}_b$---and the $\mathbf{a}_b^{*}(\boldsymbol \sigma_b, \hat{q}^{\bar\pi}_{\tau,b})$ values.   Figure \ref{fig:bootstrap_loss_illustration} shows the loss curves for the bootstrap samples (black), separated by field region.  Specifically, for a given field region $l \in \{1,2\}$, each black curve represents
\begin{equation} \label{eq:bootstrap_loss}
    \frac{1}{N_{b}} \sum_{\{j : \sigma_{b,j} \in \mathcal{P}_l\}} \mathbbm{1}(a_{b,j} \neq a_{b,j}^*(\sigma_{b,j}, \hat{q}^{\bar\pi}_{\tau_l,b}))
\end{equation}
for all $\tau \in [0,1]$ in the corresponding bootstrap sample of games.  The rugs show the optimal $\tau$-sets for each loss curve (i.e., the $\tau$ values that minimize \eqref{eq:bootstrap_loss}).  The bottom plots in Figure \ref{fig:bootstrap_loss_illustration} show weighted KDEs of the optimal $\tau$-sets for each region.\footnote{The reason we use weighted KDEs is because the sizes of the optimal $\tau$-sets vary from bootstrap sample to bootstrap sample.  For each field region, we ensure that each bootstrap sample's estimated optimal $\tau$-set has equal weight in the density estimate, regardless of size of the set.}    
 
These plots illustrate the total uncertainty on $\hat{\boldsymbol{\tau}}$, which comes from two sources.  First, for any single bootstrap sample and field region $\ell$, uncertainty arises due to the fact that multiple values of $\tau$ may minimize \eqref{eq:bootstrap_loss}.  Second, we also have uncertainty about the underlying transition probabilities, and hence the corresponding value/quantile estimates, which we account for via bootstrapping. 

The relatively large distribution for $\hat{\tau}_2$ is not surprising.  Given that there are essentially only two feasible actions and fifty states in each field region, there will inevitably be regions of the policy space that are identical.  This is particularly pronounced at the extremes of the policy space.  In each field region, there comes a point at which, no matter the value of $\tau_{\ell}$, it is impossible to have a more extreme policy.  In other words, there comes a point at which punting \textit{no matter what} (or, on the flip side, going for it no matter what) is the simply the most extreme policy, even if your latent risk preferences are ``more conservative" than another coach who always punts.  In the results section, we restrict the inference on $\tau_{\ell}$ to the interval [0.2, 0.8] due to these plateaus at the extremes of the parameter space, but note that for observed policies that lie in these extreme regions, the uncertainty intervals technically expand to the limits (either to 0 or to 1).


\section{Results} \label{sec:results}

Here we present the results of applying the methodology in Sections \ref{sec:inverse_problem} and \ref{sec:inference} to various subgroupings of fourth down decisions.  For each case, we compare the coaches' fitted risk preferences (i.e., their $\hat{\tau}_{\ell}$ values) to those fitted to the prescriptions of a win probability model.  To make this comparison, we solve \eqref{eq:multitau_exact} on the fourth down decision prescriptions from \cite{nfl4th}, hereafter referred to as the ``4th Down Bot''.  This provides a ``translation" of the risk levels underlying the 4th Down Bot to the quantile scale.  

\subsection{Risk preferences by win probability}

To estimate the $\tau$-optimal policies conditional on a given win probability range, we include this as an additional condition in the indicator function in \eqref{eq:lambda_hat} and \eqref{eq:theta_hat}.  We use the win probability estimates from \cite{nflfastR}, which employs tree-based methods trained on the same features in our state space in addition to other variables such as score differential, time remaining, and the ``Vegas line'' (i.e., the pregame point spread established by oddsmakers in Las Vegas sportsbooks), which provides a notion of the relative strength of the competing teams.\footnote{For a more complete description of this win probability model see \cite{baldwin2021nflfastr}.} Subsequently, when solving \eqref{eq:multitau_exact}, we only consider fourth down decisions that occurred in plays where the estimated win probability is in the corresponding range of interest.  The win probability ranges we consider are those given by partitioning the interval [0, 1] into twenty equally spaced bins: $[0, 0.05], (0.05, 0.1],\ldots, (0.95, 1]$.    

\begin{figure}
\centering
    \subfloat[]{
        \includegraphics[width=.71\linewidth]{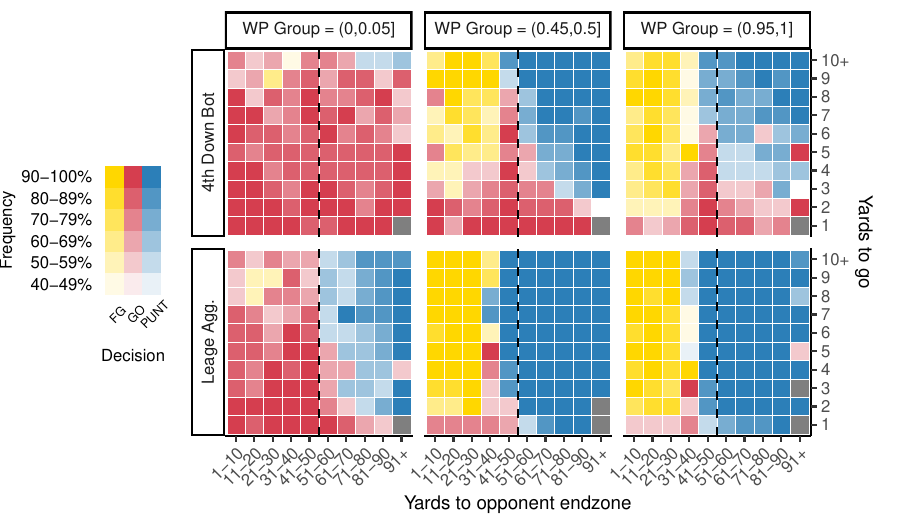}
        \label{fig:wp_decisions}
        }
    \hspace{.25em}
    \subfloat[]{
        \includegraphics[width=.25\linewidth]{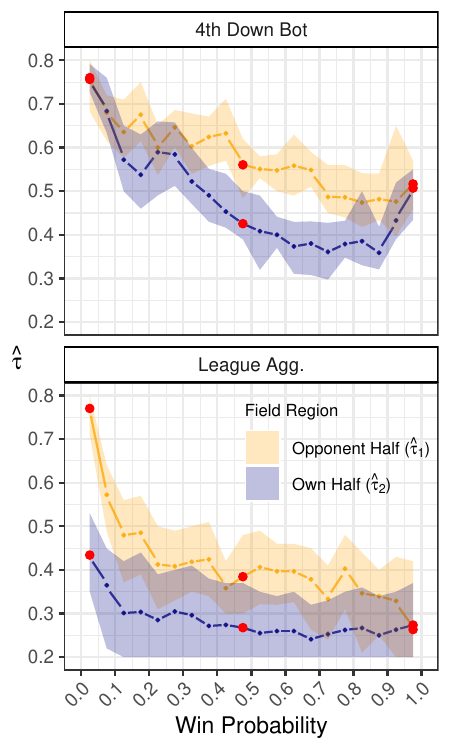}
        \label{fig:win_prob_densities}
        }
    \vspace{.75em}
    
    \subfloat[]{
        \includegraphics[width=1\linewidth]{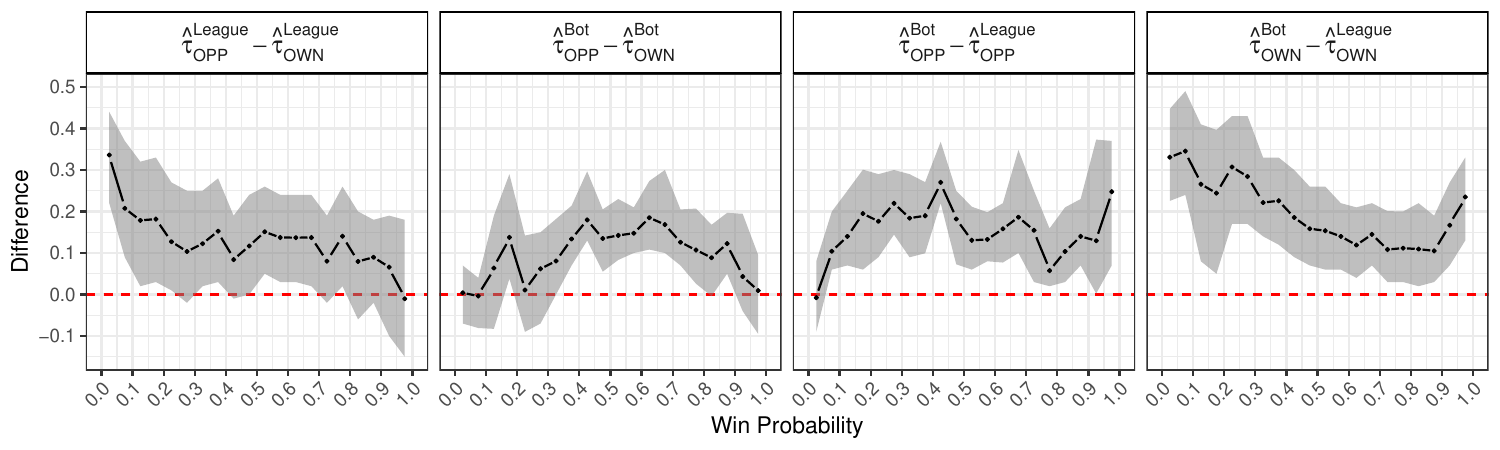}
        \label{fig:tau_hat_differences}
        }
\caption{\textbf{(a)} Top row: most frequent fourth down decision prescriptions from \cite{nfl4th} with respect to yardline and yards to first down, over the 2014-2022 NFL seasons, for three win probability bins: left - (0, 0.05]; middle - (0.45, 0.5], right - (0.95, 1].  Bottom row: most frequent observed fourth down decisions, aggregated across all coaches, for the same win probability bins.  The dashed vertical line in each plot distinguishes the Opponent half (left) from the Own half (right). All other plot features are as described in Figure \ref{fig:observed_vs_optimal}. \textbf{(b)} Estimated $\hat{\tau}$ (y-axis) for the 4th Down Bot (top) and the league in aggregate (bottom), stratified by win probability (x-axis) and half-of-field (color).  Point estimates are shown by colored dots while 95\% confidence intervals for the estimates are shown by the shaded regions.  The red dots in (b) correspond to the plots in (a). \textbf{(c)} Differences between estimated $\hat{\tau}$ values over four comparison groups of interest, with corresponding 95\% confidence intervals.}
\label{fig:win_prob_tau_hat}
\end{figure}

Figure \ref{fig:wp_decisions} shows the 4th Down Bot prescriptions (top row) and the coaches' observed decision maps (bottom row) for three of these win probability bins (left - [0, 0.05]; middle - (0.45, 0.5]; right - (0.95, 1]).  Figure \ref{fig:win_prob_densities} shows the $\hat{\tau}$ estimates (y-axis) for the 4th Down Bot (top) vs. the league in aggregate (bottom), as a function of win probability (x-axis) and stratified by field region (orange - opponent half, purple - own half).  The point estimates are shown by colored dots while 95\% confidence intervals are shown by shaded regions.  The red dots in (b) correspond to the six plots in (a).

An notable takeaway from Figure \ref{fig:win_prob_tau_hat} is that coaches appear to have different risk preferences depending on the field region in which they make a decision. This is evident from the first panel of Figure \ref{fig:tau_hat_differences}, which shows the league aggregate $\hat{\tau}_1 - \hat{\tau}_2$ values, along with corresponding 95\% confidence intervals.  Except in the (0.95, 1.0] win probability range, the differences are all positive, and the confidence intervals essentially do not contain 0 until the win probability gets to 0.8 or greater.  For low win probability fourth down situations, the average coach is clearly more willing to take risks in their opponent's half of the field than in their own half, but the magnitude of this difference dissipates as win probability increases. 

Interestingly, this same trend appears to exist for the 4th Down Bot. While the differences are not as pronounced for some win probability ranges, the second panel of Figure \ref{fig:tau_hat_differences} shows that the 4th Down Bot likewise appears more risk-tolerant in the opponent half of the field.  This surprised us; since the objective function underlying the 4th Down Bot prescriptions is identical between field regions (it is based solely on win probability regardless of game situation) we expected the ``translated'' risk tolerances of the 4th Down Bot to be identical between field regions.  The fact that they are not could be explained in part by systematic selection bias inherent in the observational data, as detected in \cite{grafstein2023correcting}. Teams that are more likely to succeed on a fourth-down attempt are more likely to go for it when given a choice, whereas teams less likely to succeed are more likely to punt or attempt a field goal.  Since the rates of these situations differ by field region (better teams face more fourth down situations in the opponent's half of the field), this could exacerbate the differences in estimated risk tolerances between the two field regions.  Additionally, the 4th Down Bot relies on a model structure of gradient boosted trees with complex interactions, which also could contribute to these differences.
  
 Finally, in both field regions and in virtually every win probability range, the risk tolerance levels estimated from the league aggregate behavior tend to be less than those of the 4th Down Bot.  This is evident from the third and fourth panels of Figure \ref{fig:tau_hat_differences}, which show $\hat{\tau}^{Bot}_1 - \hat{\tau}^{League}_1$ (opponent half differences) and $\hat{\tau}^{Bot}_2 - \hat{\tau}^{League}_2$ (own half differences), respectively.  The gaps are largest in the Own Half region with low win probability situations.\footnote{These discrepancies may be overstated to some degree, since the 4th Down Bot from \cite{nfl4th} is subject to the biases/confounding identified in \cite{grafstein2023correcting} and \cite{lopez2020bigger}.}  However, there is one subgroup that does not follow this trend; in the opponent half-of-field, when win probability is less than 0.05, the league aggregate risk preferences are commensurate with those of the 4th Down Bot.  It appears that when a team's chances of winning are very small, this is enough to spur coaches to act in accordance with the win probability objective.   

 \begin{figure}
\centering
    \subfloat[]{
        \includegraphics[width=.6\linewidth]{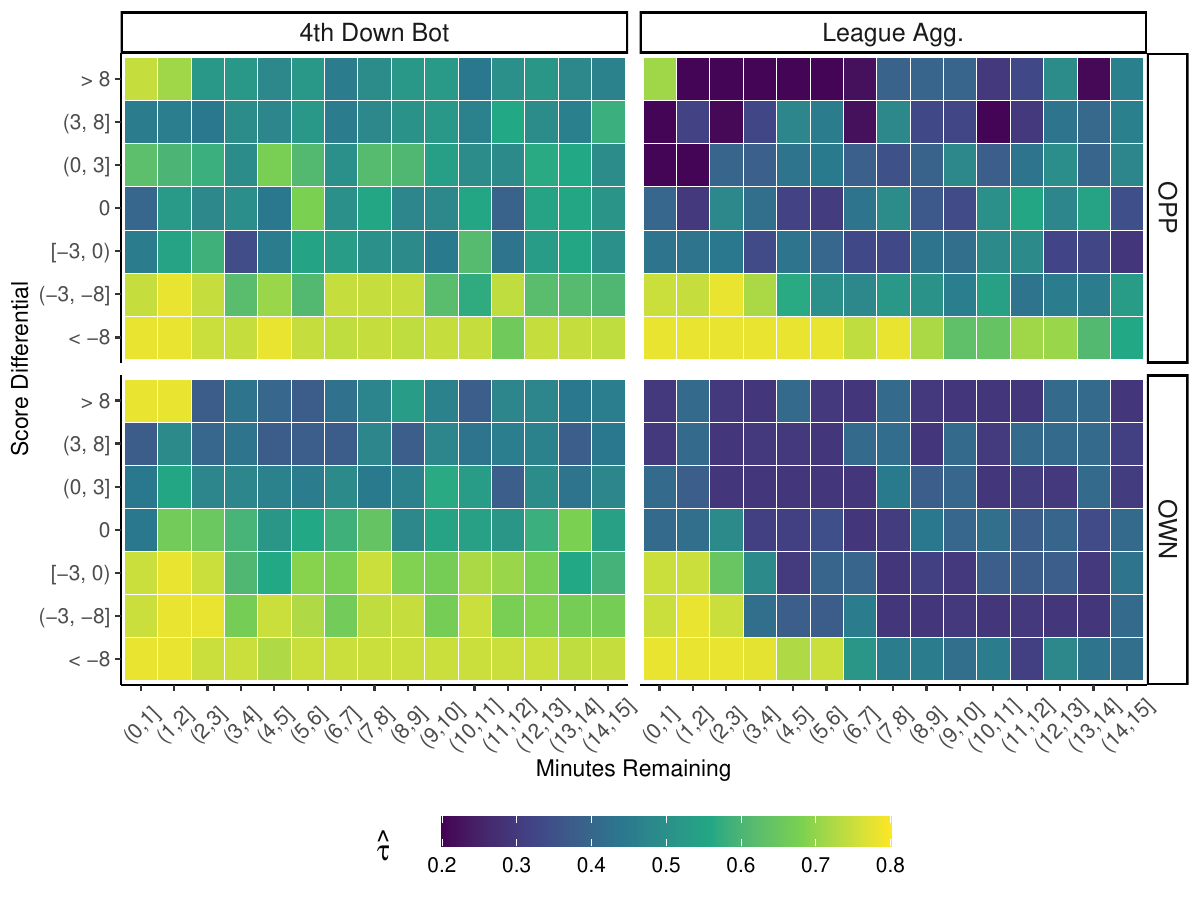}
        \label{fig:new_fig_a}
        }
    \hspace{.25em}
    \subfloat[]{
        \includegraphics[width=.3\linewidth]{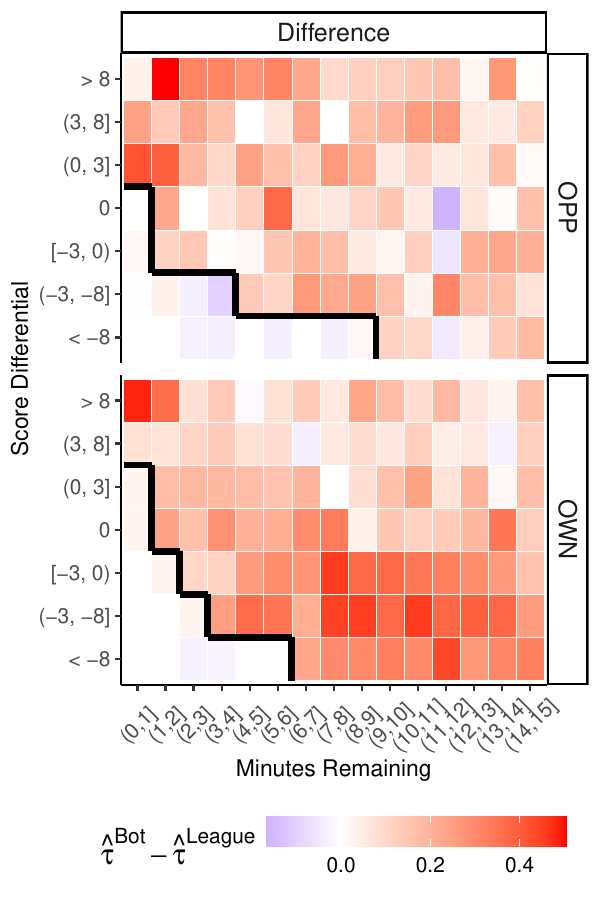}
        \label{fig:new_fig_b}
        }
\caption{\textbf{(a)} Left column: 4th Down Bot $\hat{\tau}$ estimates stratified by minutes remaining in the fourth quarter (x-axis), score differential bin (y-axis) and field region (panel rows). Right column: corresponding $\hat{\tau}$ estimates for the league aggregate fourth down behavior.  \textbf{(b)} Difference in $\hat{\tau}$ between the 4th down bot and the league average ($\hat{\tau}^{Bot} - \hat{\tau}^{League}$), stratified as in (a).  Blue regions show areas where the league average risk tolerance exceeds the 4th Down Bot's, white areas show regions where the risk tolerances align, and red areas show areas where the league average risk tolerances are less than those of the 4th Down Bot.  The black lines demarcate where coaches appear to consistently act in accordance with win probability.}
\label{fig:reference_pic}
\end{figure} 

To contextualize these findings, we investigated how the estimated risk tolerances vary when stratified by score differential and time remaining in the game, which are two of the most salient factors governing win probability, particularly late in a game.  To do this, we solve \eqref{eq:multitau_exact} on the league's fourth down decisions stratified across score differential and minutes remaining in the fourth quarter.  We binned score differential into groups based on the point values of touchdowns and field goals (e.g., score differential $\in [-3, 0)$, which encompasses situations where the team in possession needs at least a field goal to either win or tie the game).  Figure \ref{fig:new_fig_a} shows the estimated league aggregate risk tolerances stratified by these score differential bins, minutes remaining and field region.  We also included the 4th Down Bot's ``translated'' risk preferences for reference.  Figure \ref{fig:new_fig_b} shows the differences in the estimated risk tolerances between the 4th Down Bot and the league in aggregate.

Across both field regions, the majority of cells are shaded red, indicating areas where the 4th Down Bot demonstrates greater risk tolerance compared to the average coach.  This harmonizes with the findings discussed previously in relation to Figure \ref{fig:tau_hat_differences}.  However, as the fourth quarter progresses and teams fall further behind, coaches appear to become increasingly willing to make decisions based on win probability. This trend holds for both field regions.  Of particular interest are the white regions, which indicate areas where the league aggregate risk tolerances align with those of the 4th Down Bot (i.e., situations where coaches' behavior is consistent with the win probability objective).   We superimposed a black line on the plots in Figure \ref{fig:new_fig_b} demarcating where coaches appear to consistently act in accordance with win probability.

We note that there are other factors that affect the win probability of a given fourth down situation beyond those explored here (e.g., team strength and timeouts remaining).  Future research could involve studying coaches' risk sensitivities over these additional features, providing further insight into their fourth down behavior.  

\subsection{Risk preferences by quarter}

\begin{figure}
 \centering
    \subfloat[]{
        \includegraphics[width=.6\linewidth]{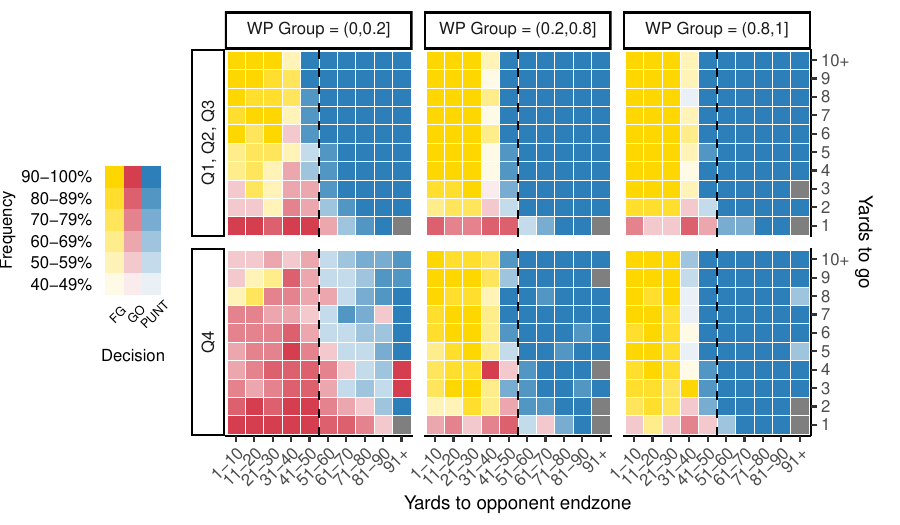}
        \label{fig:quarter_decisions}
        }
    \subfloat[]{
        \includegraphics[width=.38\linewidth]{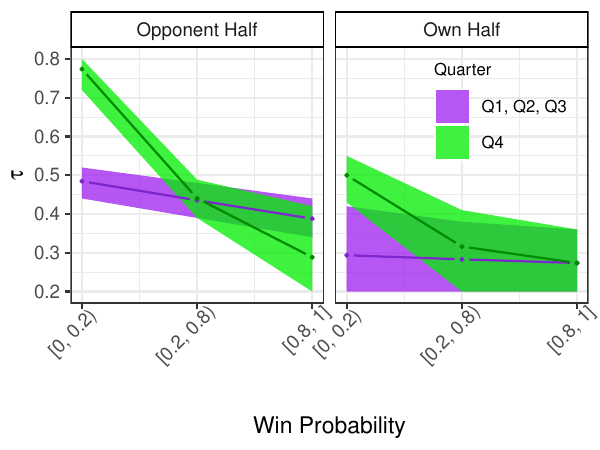}
        \label{fig:quarter_densities}
        }
\caption{\textbf{(a)} Observed fourth down decisions aggregated over the 2014-2022 NFL seasons, stratified by whether or not the decision took place in the fourth quarter (rows) and win probability group (columns). \textbf{(b)} Estimated $\hat{\tau}$ (y-axis) and corresponding 95\% confidence intervals for the league aggregate behavior under the stratification scheme shown in (a).  Purple denotes Q1, Q2, and Q3, while green denotes Q4.}
\label{fig:quarter_tau_hat}
\end{figure}

Next we explore whether coaches' risk preferences changed when making decisions in the fourth quarter, controlling for win probability.  Figure \ref{fig:quarter_decisions} depicts  coaches' observed fourth down decisions aggregated across our 9 seasons of data, stratified by whether or not the decision took place in the fourth quarter (rows) and win probability range (columns).  We used larger win probability bins for this analysis since data sparsity became an issue for many ranges in the previous win probability partition.  Specifically, in this analysis we grouped win probability into three bins: [0, 0.2), [0.2, 0.8), and [0.8, 1].  

Figure \ref{fig:quarter_densities} shows the $\hat{\tau}$ estimates (y-axis) and corresponding 95\% confidence intervals for the league aggregate behavior under the stratification scheme described above.  In this figure, purple denotes estimates for quarters 1, 2, and 3 (Q1, Q2, Q3) and green denotes estimates for quarter 4 (Q4). 

It appears that coaches' risk preferences on fourth down are essentially indistinguishable between Q4 and Q1-Q3, when in the (0.2, 0.8] and (0.8, 1] win probability ranges.  However, when the win probability is lower than 0.2, coaches are clearly much more risk-tolerant in Q4 than in Q1-Q3. Previous papers have omitted all fourth down decisions in Q4 when performing their analyses (e.g., \cite{romer2006firms, grafstein2023correcting}).  Our analysis suggests that this is only necessary in the low win probability group, for which we see drastically different risk preferences between Q1-Q3 and Q4.  As such, in the remainder of our results, we only omitted Q4 decisions when analyzing the low win probability group. 

\subsection{Risk preferences by season}

\begin{figure}
\begin{center}
        \includegraphics[width=.96\linewidth]{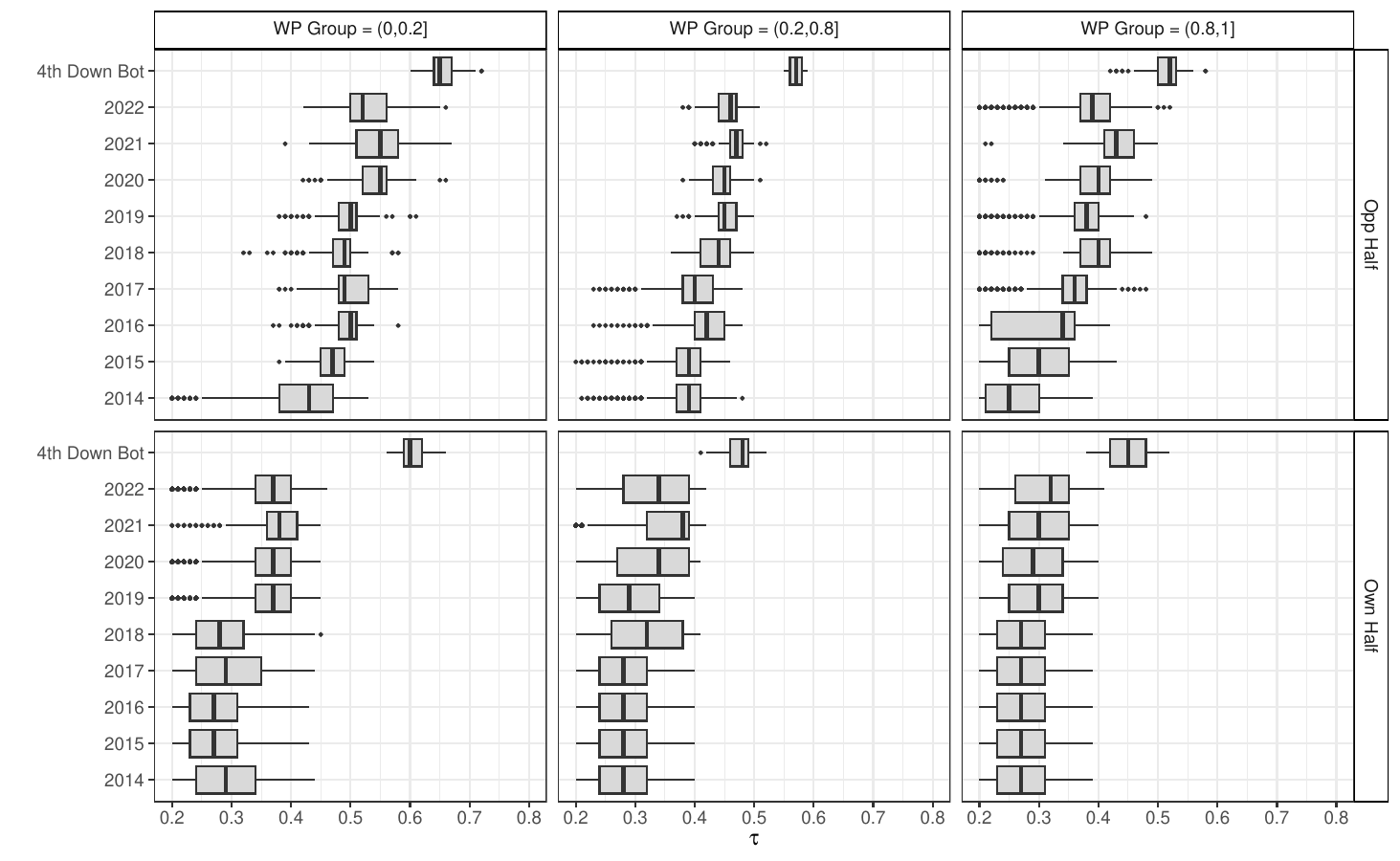}
        \caption{Weighted boxplots of 200 bootstrap solutions to \eqref{eq:multitau_exact}, stratified by win probability range (columns), field region (rows), and season (rows within plots).  The 4th Down Bot boxplots are also included for reference.}
        \label{fig:season_quantile_intervals}
\end{center}
\end{figure}

Figure \ref{fig:season_quantile_intervals} shows weighted boxplots of the 200 bootstrap solutions to \eqref{eq:multitau_exact}, stratified by win probability range (columns), field region (rows), and season (rows within plots).  The 4th Down Bot risk tolerances are also shown for reference, which are not calculated with respect to season.  It appears that risk tolerances have increased over time in every win probability range and field region combination, however the increases are more pronounced in the opponent half-of-field.  

\subsection{Risk preferences by coach and team} 

\begin{figure}
\begin{center}
        \includegraphics[width=1\linewidth]{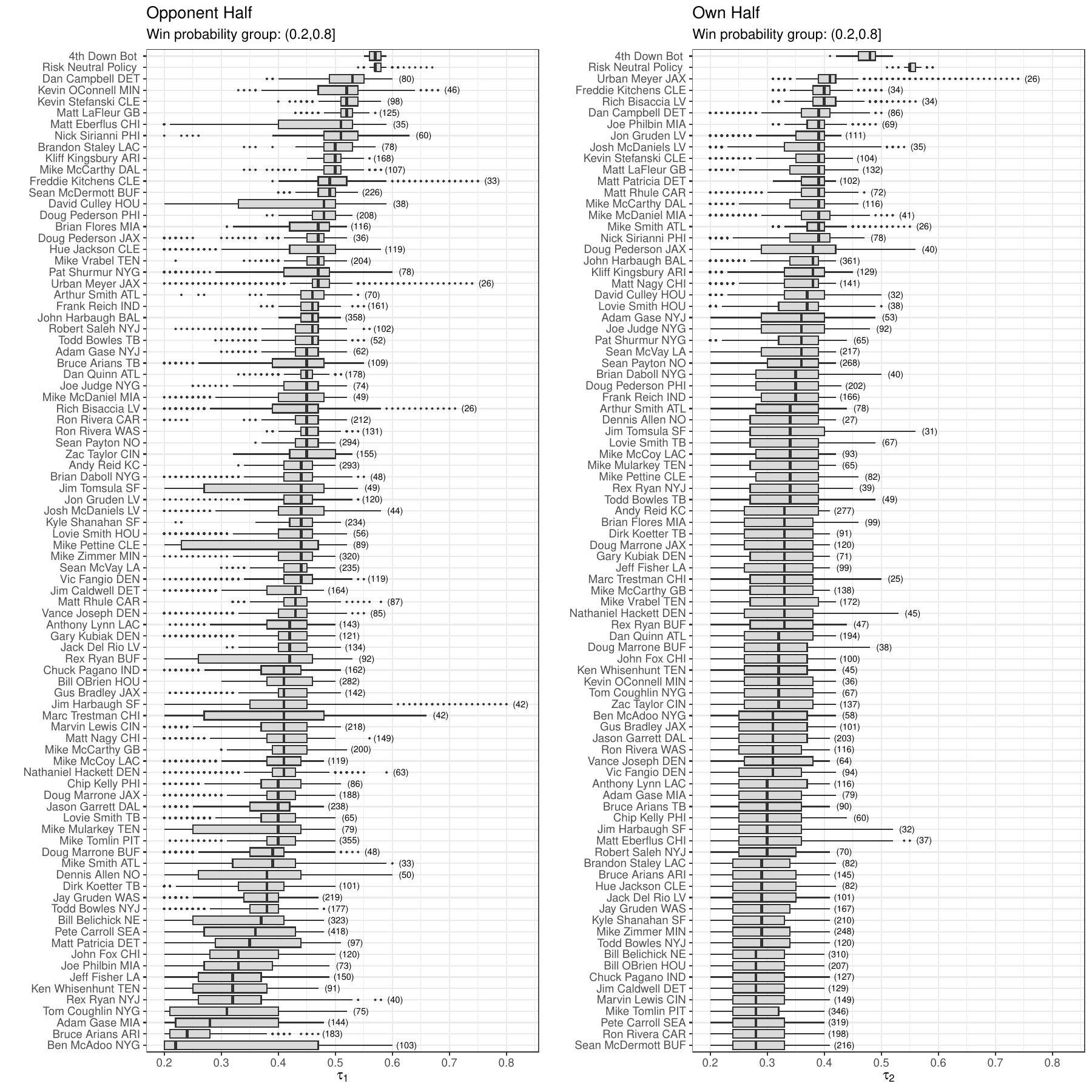}
        \caption{Weighted boxplots of 200 bootstrap solutions to \eqref{eq:multitau_exact} for decisions made in the (0.2, 0.8] win probability range, stratified by field region.  There are some coaches who appear twice in this plot (e.g., Mike McCarthy) since these coaches had at least 25 fourth down decisions with two different teams over the span of our data. In each panel, the top two boxplots correspond to the 4th Down Bot and risk-neutral policies, respectively.}
        \label{fig:coach_quantile_med}
\end{center}
\end{figure}

\begin{figure}
\begin{center}
        \includegraphics[width=.96\linewidth]{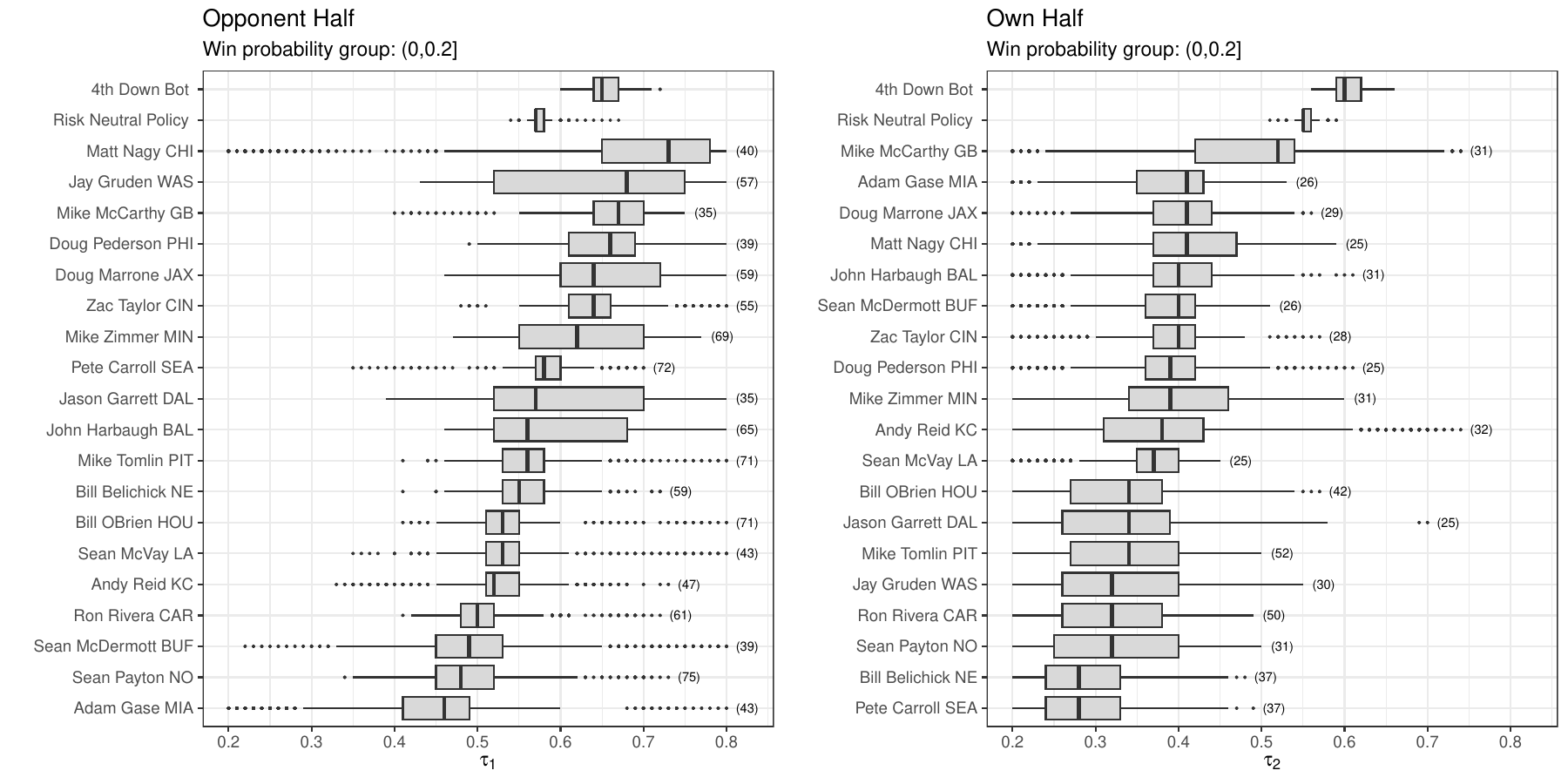}
        \caption{Weighted boxplots of 200 bootstrap solutions to \eqref{eq:multitau_exact} for coaches' decisions made in the [0, 0.2] win probability range, stratified by field region.   The number of observed decisions for each coach-team combination is shown in parentheses to the right of each boxplot.
        In each panel, the top two boxplots correspond to the 4th Down Bot and risk-neutral policies, respectively.
        }
        \label{fig:coach_quantile_low}
\end{center}
\end{figure}

\begin{figure}
\begin{center}
        \includegraphics[width=.96\linewidth]{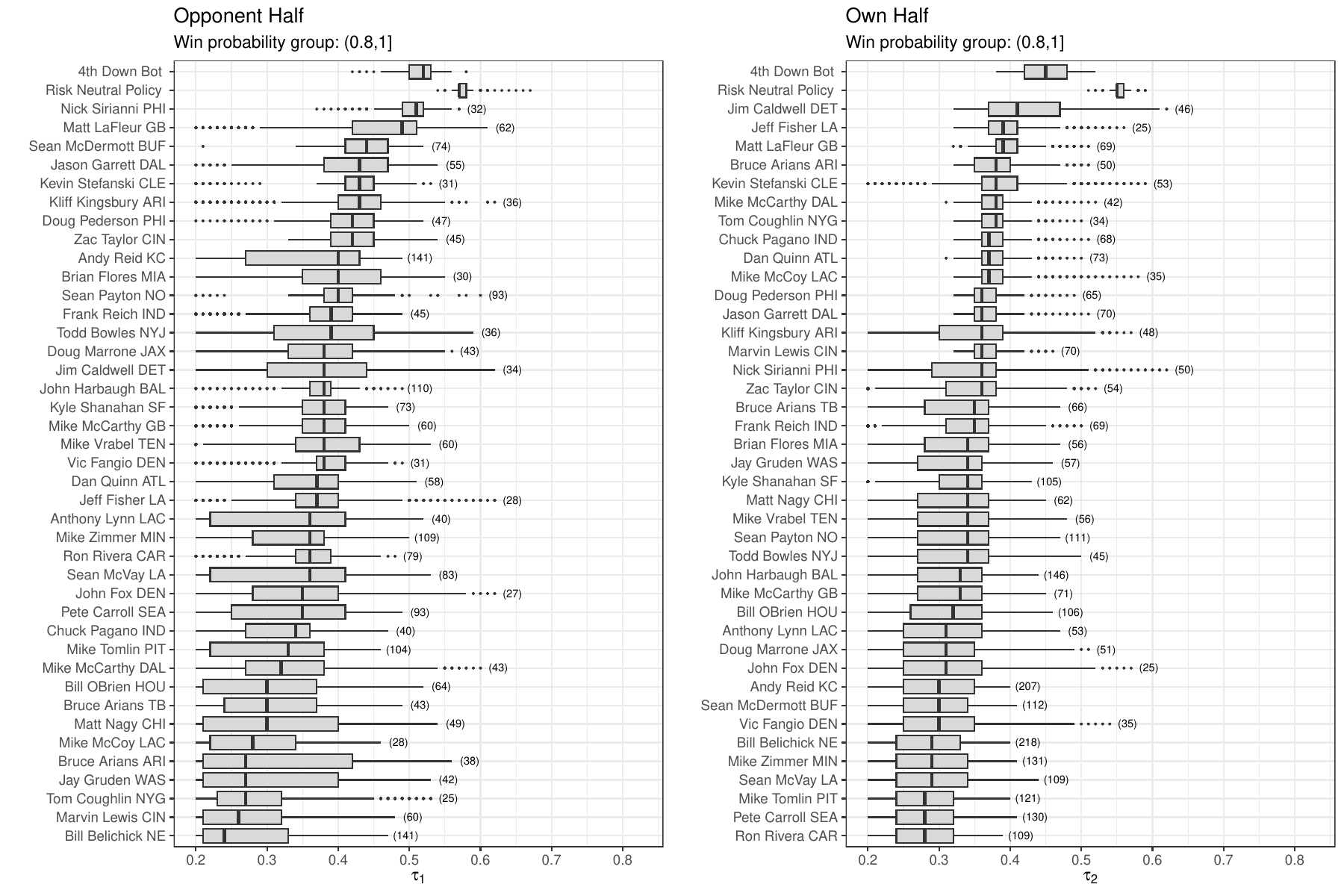}
        \caption{Weighted boxplots of 200 bootstrap solutions to \eqref{eq:multitau_exact} for decisions made in the (0.8, 1] win probability range, stratified by field region. In each panel, the top two boxplots correspond to the 4th Down Bot and risk-neutral policies, respectively.}
        \label{fig:coach_quantile_high}
\end{center}
\end{figure}

Figures \ref{fig:coach_quantile_med}, \ref{fig:coach_quantile_low}, and \ref{fig:coach_quantile_high} show weighted boxplots of the 200 bootstrapped $\hat{\tau}_{\ell}$-sets when the observed decisions have been stratified by coach-team combination and win probability range.  We only included coaches who had at least 25 observed fourth down decisions in each field region in the corresponding win probability range.  Besides the 4th Down Bot, we also estimated the risk-neutral policy and solved for the corresponding $\hat{\tau}_{\ell}$-sets for an additional point of reference in these figures.  

Several observations apparent in the league aggregate plots are also evident here.  For example, coaches are almost exclusively less risky than the 4th Down Bot and more risk-tolerant in the opponent's half-of-field.  Particularly interesting is the increased variation in estimated risk preferences across coaches in the Opponent Half region compared to the Own Half.  Coaches' behavior in their own half is generally uniform, whereas a much greater range of risk preferences are displayed in the opponent's half.  

It is also interesting to compare the coach-team boxplots with that of the risk-neutral policy.  In the Own Half region, there are no coaches whose median $\hat{\tau}_2$ is greater than that of the risk-neutral policy, across all win probability ranges.  In essence, no matter the game situation, virtually every coach is unwilling to be risky when making fourth down decisions in their own half.  On the other hand, in the Opponent Half, about half the coaches are risk-seeking (as compared to the risk-neutral policy) in the low win probability group.  In fact, the median $\hat{\tau}_1$ values for Matt Nagy, Jay Gruden, Mike McCarthy, and Doug Pedersen even exceed that of the 4th Down Bot in this win probability group.  Clearly coaches are much more willing to take risks in the opponent's half of the field than their own.    

Figures \ref{fig:coach_quantile_med}-\ref{fig:coach_quantile_high} also show that the reference boxplots are generally narrower than those of the coaches.  We suspect there are a few reasons for this.  First, in many cases the coaches' estimated risk preferences fall in underidentified areas of the model space and hence yield larger uncertainty intervals.  Second, there is likely significant variability in any given coach's risk tolerance over time; the coach's risk preferences may vary from season to season, game to game, or even decision to decision, whereas the objective functions governing the risk-neutral policy and the 4th Down Bot never vary.  Lastly, the 4th Down Bot prescriptions in each bootstrap sample are based on the win probability point estimates of \cite{nflfastR}, as opposed to refitting a win probability model for each bootstrap sample as in \cite{brill2023analytics}, which may lead to narrower confidence intervals.

\subsection{Relationship to fourth down performance}

To investigate the relationship between the coaches' estimated risk preferences and their performance on fourth down, we regressed the coaches' estimated risk preferences on their average \textit{points gained} on fourth down plays \citep{chan2021points}.  Specifically, for coach $i$, in season $j$, in win probability range $k$, and field region $\ell$, we computed: 
\begin{align}
\text{Avg Points Gained}_{ijk\ell} = \frac{1}{N_{ijk\ell}}\sum_{g \in \mathcal{G}_{ij}} \sum_{t \in \mathcal{S}^4_{gk\ell}} (\hat{v}^{\bar\pi}(s_{t+1}) - \hat{v}^{\bar\pi}(s_t)),
\end{align}
where $\mathcal{G}_{ij}$ is coach $i$'s set of games in season $j$, $\mathcal{S}^4_{gk\ell}$ is the set of play indices in game $g$ which occur in win probability range $k$ and field region $\ell$, and $N_{ijk\ell}$ is the total number of play indices which satisfy these criteria.  We also wanted to include a covariate in the model to control for the team's overall offensive strength for the corresponding season.  To do this, we gathered game-by-game Elo ratings of every NFL team and computed the team averages over each season in our data, yielding a unique $\text{Elo}_{ij}$ value for each coach/season combination in our data.\footnote{We used the Elo values from FiveThirtyEight.com, which site has since been archived.  These data are still publicly available however at \url{https://projects.fivethirtyeight.com/nfl-api/nfl_elo.csv}.}  

Equipped with these quantities, we model coach $i$’s average points gained on fourth down in season $j$, in win probability range $k$, in field region $\ell$ as
\begin{align}
\text{Avg Points Gained}_{ijk\ell} &= \beta_0 + \beta_1 \times \hat{\tau}_{ijk\ell} +\beta_2 \times \text{Elo}_{ij}+\beta_3 \times \mathrm{I}\left(\ell=\text{Own}\right) + \epsilon_{ijk\ell} \nonumber \\ 
 \epsilon_{ijk\ell} &\sim N(0, \sigma^2) \label{eq:reg_model}
\end{align}
where $\hat{\tau}_{ijk\ell}$ represents the coaches' estimated risk preferences, likewise stratified by season, win probability group, and field region.  We excluded all averages that were based on less than 25 fourth down decisions in the analysis.  The regression output is shown in Table \ref{table:regression_table}.

\begin{table}
\centering 
\footnotesize
\begin{tabular}{@{\extracolsep{5pt}}llrl} 

\\[-1.8ex]\hline 
\hline \\[-1.8ex] 
\\[-1.8ex] \multicolumn{2}{l}{Parameter (Covariate)} & Estimate (Std. Err.) & Partial R$^2$ \\ 
\hline \\[-1.8ex]
\multicolumn{2}{l}{$\beta_0$ (Intercept)} & $-$0.322$^{***}$ (0.058) & NA\\ 
  & \\ 
\multicolumn{2}{l}{$\beta_1$ ($\hat{\tau}$)} & 0.769$^{***}$ (0.138) & 0.048\\ 
  & \\ 
 \multicolumn{2}{l}{$\beta_2$ (Average Elo rating by team and season)} & 0.036$^{***}$ (0.010) & 0.020\\ 
  & \\ 
 \multicolumn{2}{l}{$\beta_3$ (Indicator for field region in Own Half)} & 0.079$^{***}$ (0.022) & 0.021 \\ 
\hline \\[-1.8ex] 
Observations: 622 & & &\\ 
R$^{2}$:  0.059 & & &\\ 
Adjusted R$^{2}$:  0.054 & & &\\ 
Residual Std. Error:  0.237 (df = 618)& & &\\ 
F Statistic: 12.860$^{***}$ (df = 3; 618) & & &\\ 
\hline 
\hline \\[-1.8ex] 
& \multicolumn{3}{r}{\textit{Note:} $^{***}$p$<$0.001}  \\ 
\end{tabular} 
\caption{Regression output for the model in \eqref{eq:reg_model}.} \label{table:regression_table}
\end{table} 

While the signal-to-noise ratios are low (e.g., the partial $R^2$ value for each covariate is less than 0.05), each covariate has a statistically significant linear association with average points gained on fourth down.  In particular, we observe that average points gained is positively associated with $\hat{\tau}$ (controlling for team strength and half-of-field), suggesting that a coach's excessive risk aversion on fourth down is negatively associated with team performance, which is consistent with prior research \citep{yam2018fourthdown}. 

\section{Conclusion}

The inverse optimization framework we propose in this study offers a unique approach to contextualize and understand coaches' decision-making on fourth down.  By inferring the actual optimization problem that underlies a coach's behavior, we not only are able to predict their future behavior, but we can actually learn about \textit{why} they make the decisions that they do.  We learn directly about their risk preferences and values, which the coaches themselves may not be able to articulate.  This understanding could help analysts and coaches alike in comprehending the rationale behind specific decision-making behaviors and could better facilitate behavioral change.  These methods could likewise be applied in other domains to learn about a decision maker's risk preferences.  

The study also provides insights into the risk profiles and decision-making behavior of NFL coaches in various fourth down situations.  We find differences in risk preferences based on field region, with coaches displaying higher risk tolerances when making decisions in the opponent's half of the field. Furthermore, our analysis tracks the evolution of risk tolerances over time, revealing an overall increase in risk tolerance for the league and a more pronounced increase in the opponent's half of the field.  In harmony with previous research, we find that coaches' risk tolerances generally tend to be lower than what statistical models prescribe based on win probability estimates, indicating excessive risk aversion in most game situations.  

To maintain computational feasibility and address data sparsity concerns, our analysis excludes other variables such as score differential, time remaining, and timeouts remaining from the state space. Instead, we stratify subsequent analyses by win probability, which serves as a univariate proxy for the mentioned variables. We also acknowledge that including team-specific transition probabilities would provide additional insights; however, due to data sparsity and computational constraints, we focus on league-wide estimates in this study. Future research could explore ways to incorporate team-specific transition probabilities for a more granular understanding of coaches' decision-making behaviors.

\begin{acks}[Acknowledgments]
The authors would like to thank the associate editor and anonymous referees for their constructive comments that improved the quality of this paper.
\end{acks}
%


\begin{supplement}
\stitle{Supplement A}

\sdescription{Supplement A includes Appendices A, B, and C.  Appendix A includes details on the value function derivation of \cite{chan2021points}.  Appendix B includes details on the construction of an approximately optimal size-two partition of the state space.  Appendix C includes three additional figures:
\begin{itemize}
    \item Figure \ref{fig:partition_fig} depicts four different partitions of $\mathcal{S}^4$ and the corresponding loss from \eqref{eq:multitau2} for each partition.
    \item Figure \ref{fig:league_aggregate_solution} depicts all fourth down decisions in our data, and the corresponding loss function when applying our inverse optimization method. 
    \item Figure \ref{fig:full_quantile_policy} illustrates the $\tau$-optimal policies for all values of $\tau \in \{0.20, 0.21, \ldots, 0.80\}$, which is the set over which we make inference on $\tau$. 
\end{itemize}
}
\end{supplement}


\bibliographystyle{imsart-nameyear} 
\bibliography{bibliography.bib}       

\newpage
\begin{appendix} 
\counterwithin{figure}{section}

\section{Value Function Derivation} \label{sec:value_func_derivation}

Here we provide the result of the value function derivation from \cite{chan2021points} and refer the reader to their paper for a detailed interpretation and the corresponding proofs.

The transition dynamics for the Markov reward process induced by $\bar\pi$ (described in Section \ref{sec:decision_model}) can be defined in matrix form as 
\begin{equation}
    \mathbf{P}^{\bar\pi} = \left[\begin{array}{ll}
\mathbf{P}^{\bar\pi}_{\text{A} \text{A}} & \mathbf{P}^{\bar\pi}_{\text{A}\text{B}} \\
\mathbf{P}^{\bar\pi}_{\text{B}\text{A}} & \mathbf{P}^{\bar\pi}_{\text{B}\text{B}}
\end{array}\right], \label{eq:TPM}
\end{equation}
where submatrices $\mathbf{P}^{\bar\pi}_{\text{A}\text{A}}$ and $\mathbf{P}^{\bar\pi}_{\text{B}\text{B}}$ correspond to transitions where team $\text{A}$ and team $\text{B}$ retain possession of the ball, respectively. The other two submatrices correspond to changes of possession.  Under the setting of identical teams, we have that $\mathbf{P}^{\bar\pi}_{AA} = \mathbf{P}^{\bar\pi}_{BB}$ and $\mathbf{P}^{\bar\pi}_{AB} = \mathbf{P}^{\bar\pi}_{BA}$ in the matrix representation of transition probabilities given in \eqref{eq:TPM}.  

Within this context, the relevant result from \cite{chan2021points} is that the solution to \eqref{eq:value_func_approx} is given by the following equation:
\begin{align}
     \mathbf{v}^{\bar\pi}_{A} &= \mathbf{r}_{A}\left(\mathbf{I}-\mathbf{P}^{\bar\pi}_{A A}+\mathbf{P}^{\bar\pi}_{A B}\right)^{-1} \label{eq:vector_value_func}
\end{align}
where $\mathbf{v}^{\bar\pi}_A$ is the vector of values across all states in $\mathcal{S}_A$, denoting all states where team A is in possession of the football, and where $\mathbf{r}_\text{A}$ is the corresponding vector of rewards (which is 0 for all states except the three scoring states).  The reward vector for team B is  the negative of team A's reward vector, $\mathbf{r}_\text{B} = -\mathbf{r}_\text{A}$, hence $\mathbf{v}^{\bar\pi}_{B} = -\mathbf{v}^{\bar\pi}_A$ by symmetry.  

\begin{figure}[H] 
\begin{center}
\includegraphics[width=1\textwidth]{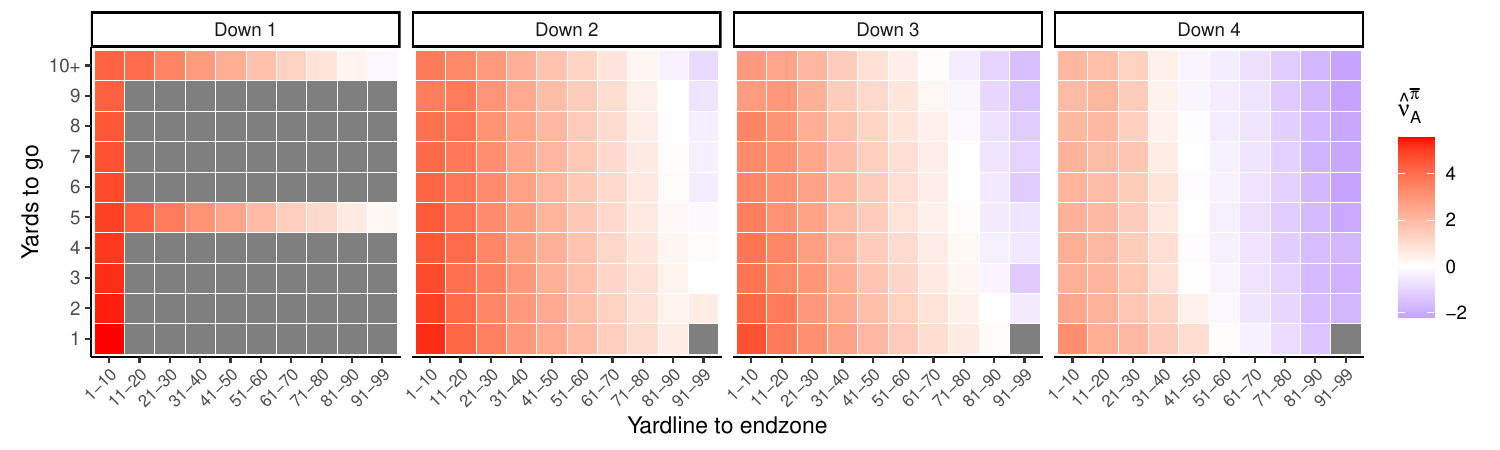}
\end{center}
\caption{Estimated values of $\mathbf{v}^{\bar\pi}_A$ for all states $s\in \mathcal{S}^{\text{play}}_A$.} \label{fig:value_function}
\end{figure}

We obtain an estimate of $\mathbf{v}^{\bar\pi}_{A}$ by plugging the empirical probabilities given by \eqref{eq:theta_hat} into $\mathbf{P}^{\bar\pi}_{A A}$ and $\mathbf{P}^{\bar\pi}_{A B}$ for all state pairs $(s, s') \in \{\mathcal{S}_A \times \mathcal{S}\}$ and solving \eqref{eq:vector_value_func}.  Figure \ref{fig:value_function} shows the estimated values $\hat{\mathbf{v}}^{\bar\pi}_A$ for all states $s\in \mathcal{S}^{\text{play}}_A$.  The trends are intuitive; a state's value decreases as any of down, yardline, or yards to go increases.
%


\newpage

\section{Constructing an approximately optimal size two partition of \texorpdfstring{$\mathcal{S}^4$}{Lg}} \label{sec:partition_appendix}

Solving \eqref{eq:multitau2} for $L=2$ is difficult given the massive number of unique qualifying partitions.  We can, however, solve \eqref{eq:multitau2} in the singleton partition setting (i.e., the parition given by $\{\{\sigma\} ; \sigma \in \mathcal{S}^4\}$) which provides a lower bound on the minimum loss irrespective of partition size.  We can then find an approximate optimal partitioning given $L=2$; as long as the resulting loss is reasonably close to the lower bound, the approximate solution can be used.  

\begin{figure}[H] 
\begin{center}
\includegraphics[width=1\textwidth]{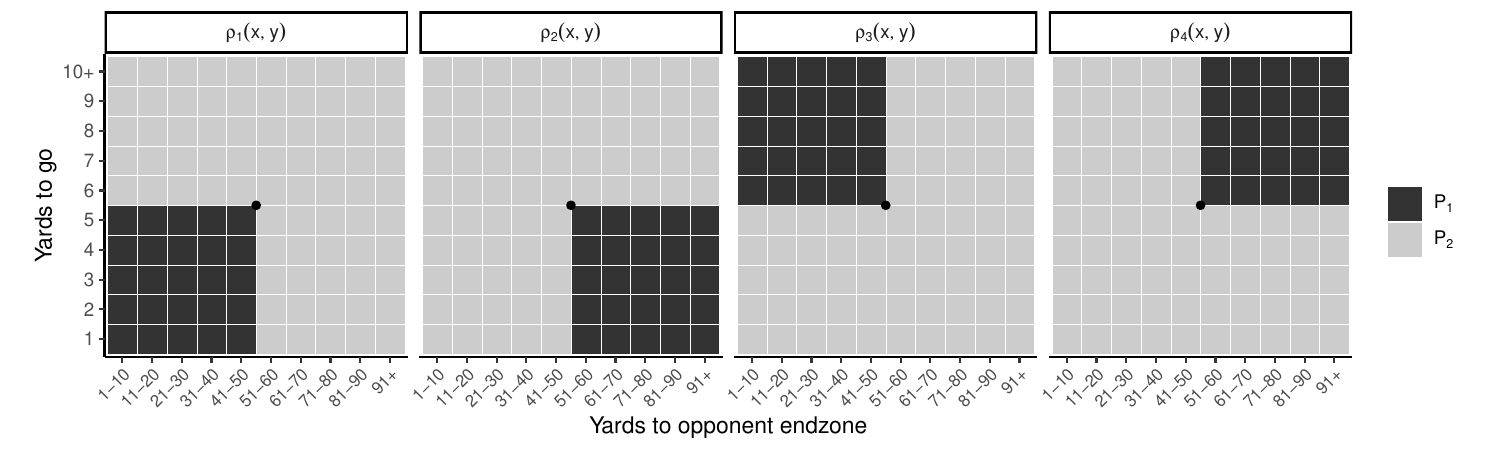}
\end{center}
\caption{Partition structure for $x = $ 51-60 and $y = $ 6.  In each plot, the black cells comprise $\mathcal{P}_1$ and the gray cells comprise $\mathcal{P}_2$ (i.e., $\mathcal{P}_1^c$).} \label{fig:partition_visual}
\end{figure}

The vast majority of size-2 partitions of the fourth down state space do not represent realistic boundaries over which coaches' risk preferences would change.  We therefore approximate \eqref{eq:multitau2} under $L=2$ by considering a subset of the partition space with a plausible form.   Let $\rho = \{\mathcal{P}_1, \ldots, \mathcal{P}_L\}$ denote a partition of $\mathcal{S}^4$. For any given fourth down state (which can be represented as a yardline/yards-to-go combination), we create four size-2 partitions as follows: 
\begin{align}
\rho_1(x,y) &= \left\{\mathcal{P}_1 = \{\sigma \in \mathcal{S}^4: X(\sigma) < x ~~\&~~ Y(\sigma) < y\}, \mathcal{P}_1^c \right\} \nonumber \\ 
\rho_2(x,y) &= \left\{\mathcal{P}_1 = \{\sigma \in \mathcal{S}^4: X(\sigma) < x ~~\&~~ Y(\sigma) \geq y\}, \mathcal{P}_1^c \right\} \nonumber \\ 
\rho_3(x,y) &= \left\{\mathcal{P}_1 = \{\sigma \in \mathcal{S}^4: X(\sigma) \geq x ~~\&~~ Y(\sigma) < y\}, \mathcal{P}_1^c \right\} \nonumber \\ 
\rho_4(x,y) &= \left\{\mathcal{P}_1 = \{\sigma \in \mathcal{S}^4: X(\sigma) \geq x ~~\&~~ Y(\sigma) \geq y\}, \mathcal{P}_1^c \right\} \nonumber 
\end{align}
where $(x, y)$ denotes a yardline/yards-to-go pair, $X(\sigma)$ returns the yardline group for state $\sigma$, $Y(\sigma)$ returns the yards-to-go for $\sigma$, and $\mathcal{P}_1^c$ is the complement of $\mathcal{P}_1$ (with $\mathcal{S}^4$ as the sample space).  Figure \ref{fig:partition_visual} illustrates this partition structure for $x = $ 51-60 and $y = $ 6.

Considering all $(x, y)$ combinations in $\mathcal{S}^4$ jointly, we end up with a total of 344 partitions which adhere to this criteria, which is easy to solve over.  The partition with the minimum loss is given by breaking the state space at the 40 yardline, the minimum loss being 0.1499938.  Second best is the partition given by breaking the state space at the 50 yardline with the minimum loss being 0.1500247.  Since these values are virtually identical, we chose the partition that separates states by own half vs opponent half, for interpretability reasons.  


\section{Additional Figures} \label{sec:additional_figs}

\begin{figure}[H]
 \centering
    \subfloat[]{
        \includegraphics[width=.5\linewidth]{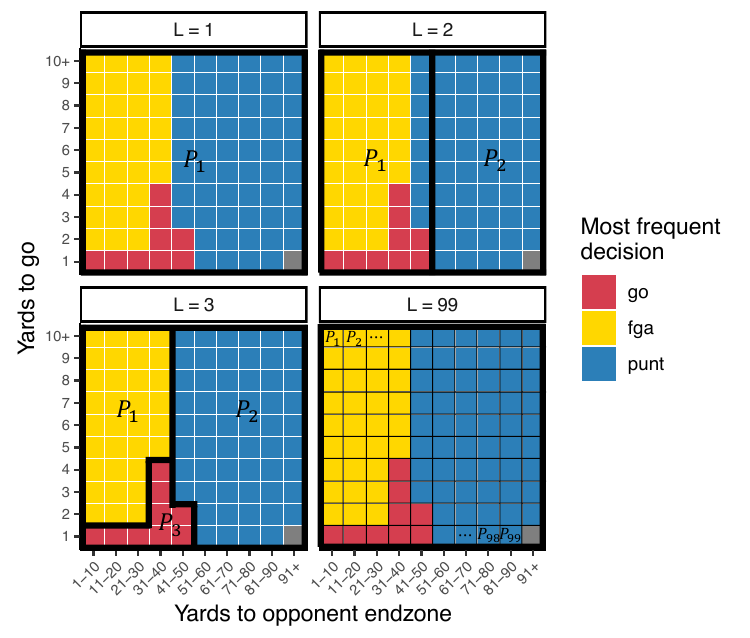}
        \label{fig:partition_a}
        }
    \subfloat[]{
        \includegraphics[width=.5\linewidth]{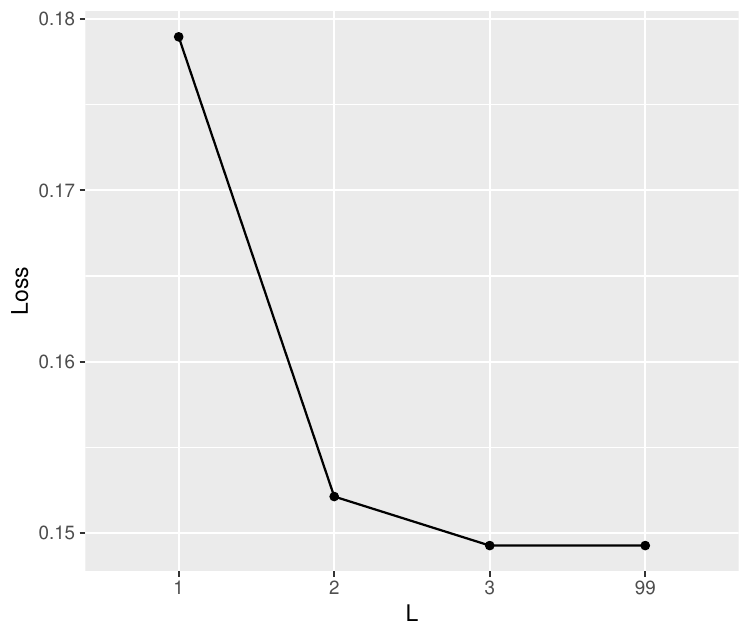}
        \label{fig:partition_loss}
        }
\caption{\textbf{(a)} Four different partitions of $\mathcal{S}^4$, of sizes $L = 1, 2,3,$ and 99.  Upper left: trivial partition. Upper right: two subset partition created by cutting the state space by the 50 yardline. 
 Lower left: optimal three subset partition with respect to \eqref{eq:multitau2}. Lower right: singleton partition where each state is its own subset. \textbf{(b)} Loss associated with solving \eqref{eq:multitau2} given the partitions shown in (a). The minimized losses for $L = 3$ and $L = 99$ (which is the most complex model possible) are identical.}
\label{fig:partition_fig}
\end{figure}

 \begin{figure}[H]
  \centering
    \subfloat[]{
        \includegraphics[width=.5\linewidth]{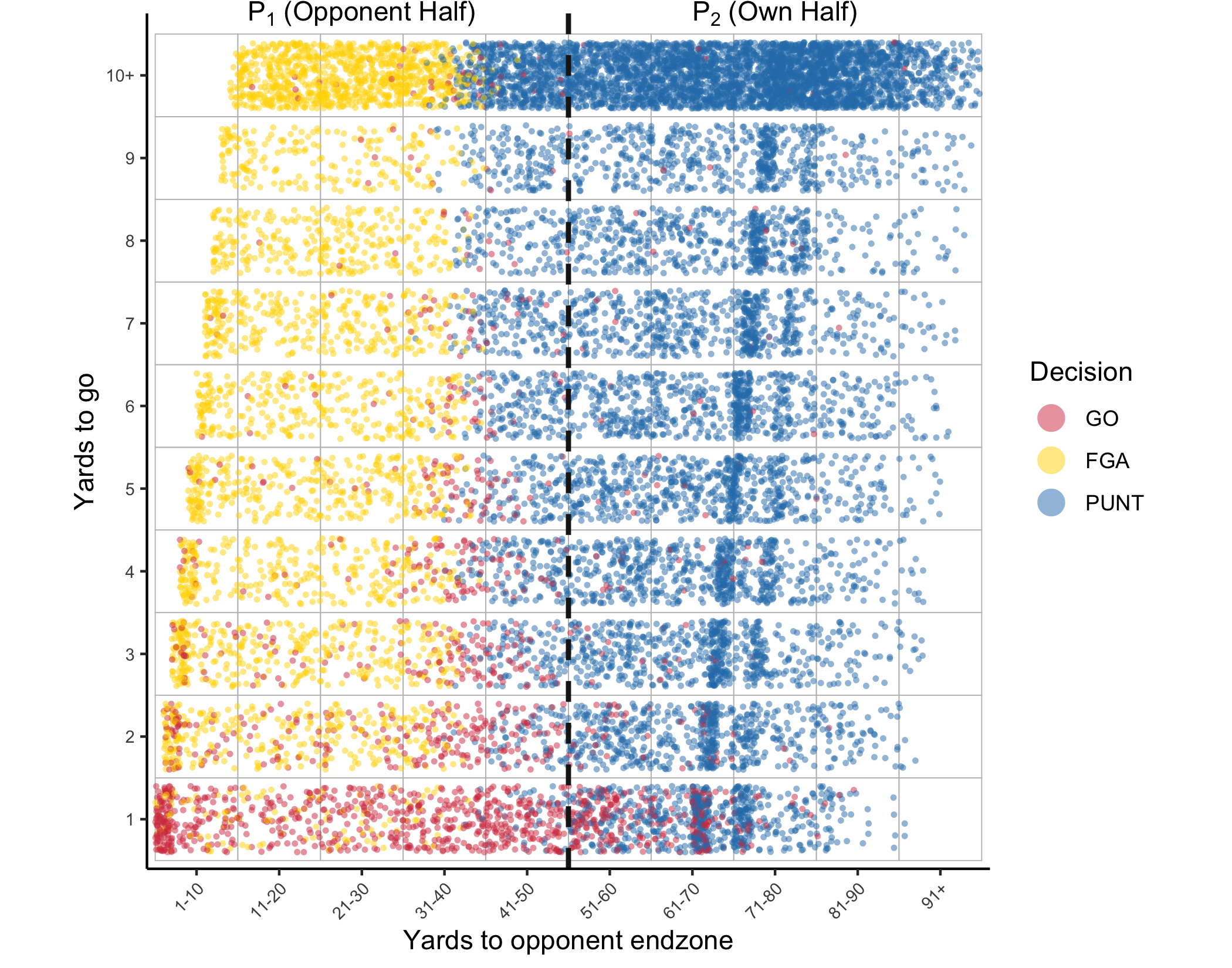}
        \label{fig:obs_decisions}
        }
    \subfloat[]{
        \includegraphics[width=.5\linewidth]{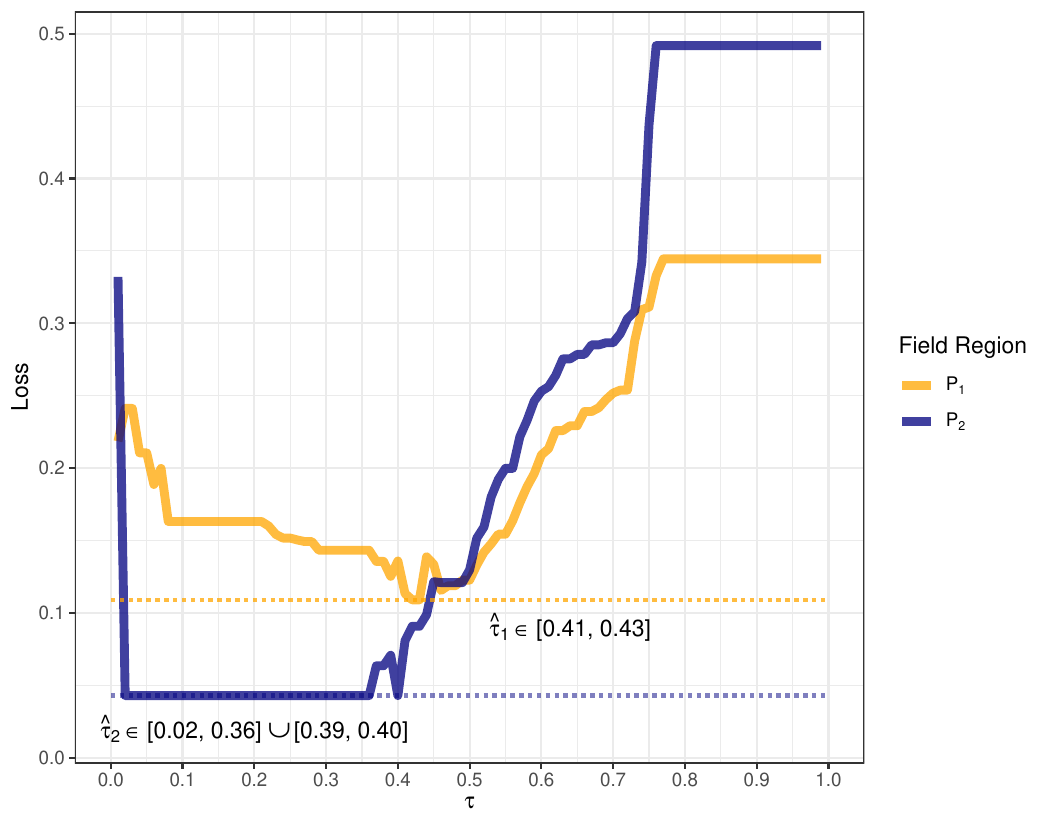}
        \label{fig:league_aggregate_loss_curves}
        }
\caption{(a) All observed 4th down decisions $\mathbf{a}$, from 2014 to 2022.  The dashed line separates $\mathcal{P}_1$ (Opponent half) and $\mathcal{P}_2$ (Own half). (b) Loss curves corresponding to the decisions $\mathbf{a} \in \mathcal{P}_1$ (orange) and $\mathbf{a} \in \mathcal{P}_2$ (purple).  The dashed horizontal lines denote the minimum loss values for each field region.  The $\tau$ values for which the loss is minimized for each field region are shown in text.  In both field regions, there are multiple $\tau$'s which satisfy \eqref{eq:multitau_exact}.}
\label{fig:league_aggregate_solution}
\end{figure}

\begin{figure}[H]
\begin{center}
\includegraphics[trim = 0in 1.15in 0in 1.1in, clip, width=1\textwidth]{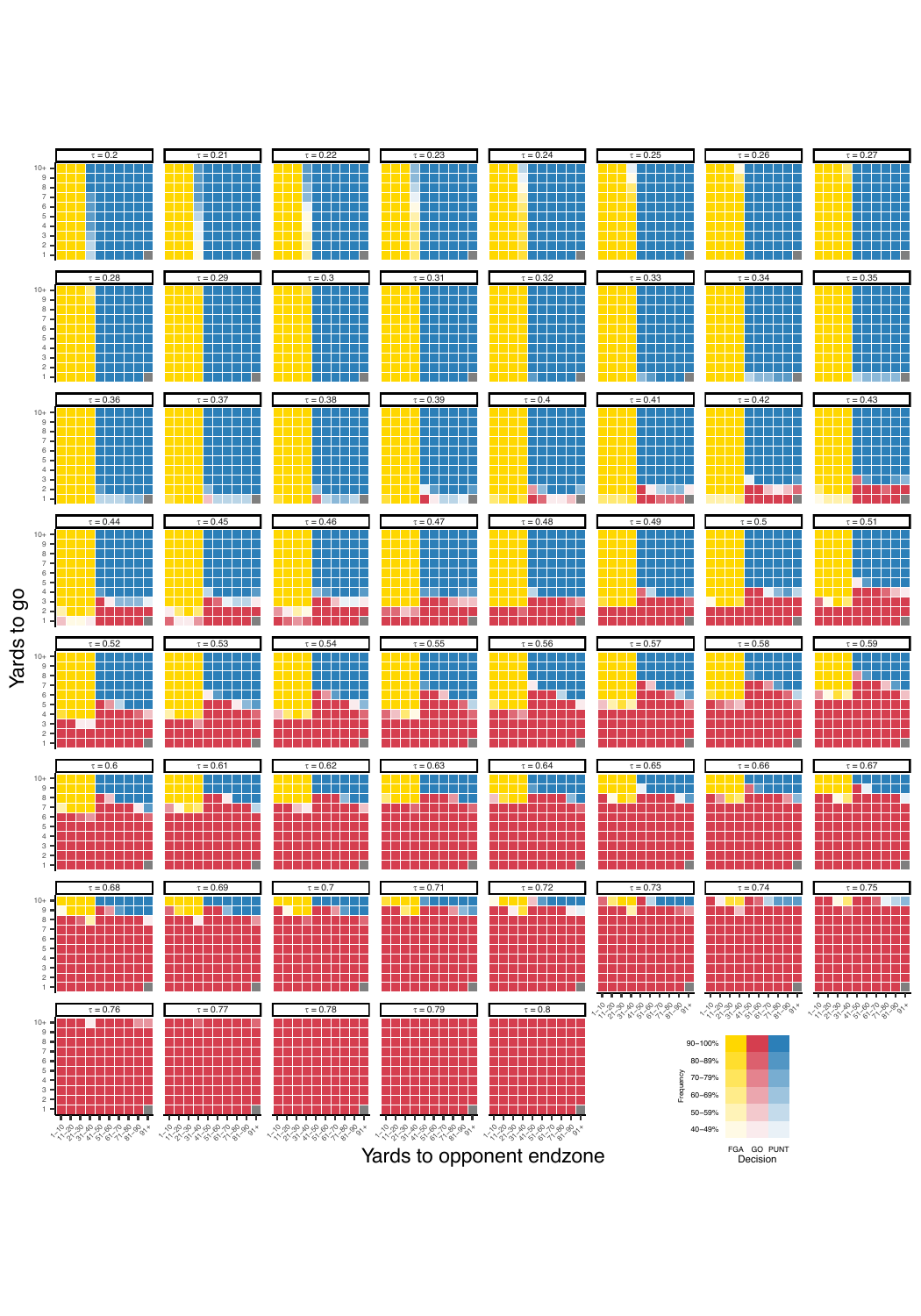}
\label{fig:full_quantile_map}
\end{center}
\caption{$\tau$-optimal policies  as given by \eqref{eq:astar} for all values of $\tau \in \{0.20, 0.21, \ldots, 0.80\}$, which is the set over which we make inference on $\tau$.  Color saturation in a given state $\sigma$ corresponds to the $a_b^*(\sigma,\hat{q}^{\bar\pi}_{\tau,b})$ frequencies over 200 bootstrap solutions.  The decision boundary between punting and field goal attempts appears to be fairly robust across bootstrap samples, while the boundary between going for it and the two kicking decisions tends to vary from bootstrap sample to bootstrap sample.} \label{fig:full_quantile_policy}
\end{figure}

\end{appendix}

\end{document}